\documentclass[
  journal=largetwo,
  manuscript=article-type,
  year=2024,
  volume=37,
]{cup-journal}

\usepackage{graphicx}%
\usepackage{multirow}%
\usepackage{amsmath,amssymb,amsfonts}%
\usepackage{amsthm}%
\usepackage{mathrsfs}%
\usepackage[title]{appendix}%
\usepackage{xcolor}%
\usepackage{textcomp}%
\usepackage{manyfoot}%
\usepackage{booktabs}%
\usepackage{algorithm}%
\usepackage{algorithmicx}%
\usepackage{algpseudocode}%
\usepackage{listings}%
\usepackage{float}
\usepackage{booktabs}
\usepackage{subcaption}
\usepackage{lineno}
\usepackage{tabularx}
\usepackage[nopatch]{microtype}
\usepackage[table]{xcolor}
\usepackage{placeins}
\usepackage{float}

\title[Title]{Probing interplanetary scintillation using broadband VLBI}

\author{A. Jaradat}
\affiliation{School of Natural Sciences, University of Tasmania, Private Bag 37, Hobart, 7005, Tasmania, Australia}
\email[A. Jaradat]{Ahmad.Jaradat@utas.edu.au}

\author{G. Molera Calv\'{e}s}
\affiliation{School of Natural Sciences, University of Tasmania, Private Bag 37, Hobart, 7005, Tasmania, Australia}

\author{J. Edwards}
\affiliation{School of Natural Sciences, University of Tasmania, Private Bag 37, Hobart, 7005, Tasmania, Australia}
\alsoaffiliation{CSIRO Space and Astronomy, Epping, 1710, NSW, Australia.}

\author{S. Ellingsen}
\affiliation{School of Natural Sciences, University of Tasmania, Private Bag 37, Hobart, 7005, Tasmania, Australia}
\alsoaffiliation{International Centre for Radio Astronomy Research, The University of
Western Australia, Crawley, 6009, WA, Australia.}

\author{T. McCarthy}
\affiliation{School of Natural Sciences, University of Tasmania, Private Bag 37, Hobart, 7005, Tasmania, Australia}

\author{J. Morgan}
\affiliation{CSIRO Space and Astronomy, Bentley, 6102, WA, Australia}


\begin{document}

\begin{abstract}
Advancements in VLBI instrumentation, driven by the geodetic community's goal of achieving positioning accuracy of 1 mm and stability of 0.1 mm/y, have led to the development of new broadband systems. Here, we assess the potential of these new capabilities for space weather monitoring. These enhanced VLBI capabilities were used to investigate interplanetary scintillation (IPS), a phenomenon caused by the scattering of radio waves due to density irregularities in the solar wind. 
Compact radio sources near the Sun were observed using the AuScope VLBI array in Australia, which consists of 12-meter telescopes at Hobart, Katherine, and Yarragadee. The baseline lengths between these telescopes are approximately 3400 km (Hobart–Katherine), 3200 km (Hobart–Yarragadee), and 2400 km (Katherine–Yarragadee). The observations covered solar elongations from 6.5$^\circ$ to 11.3$^\circ$ and frequencies between 3 and 13 GHz.
The study focused on phase scintillation as an indicator of turbulence in the solar wind. As the solar elongation decreased, we observed an increase in the phase scintillation index, consistent with theoretical models. Importantly, the broadband system also detected IPS using relatively weak radio sources. Additionally, the phase scintillation increased with baseline length, in agreement with Kolmogorov turbulence with an index of 11/3.
These findings demonstrate the effectiveness of geodetic broadband VLBI in capturing detailed features of the solar wind. This capability enables continuous space weather monitoring and advances our understanding of solar and interplanetary dynamics.
\end{abstract}


\section{Introduction}\label{sec1}

Understanding the solar wind—its density, velocity, temperature, and magnetic field—is crucial for heliophysics and space weather studies. \textit{In-situ} measurements made by spacecraft traversing the solar wind have provided precise and high-resolution data that are invaluable for this purpose. Spacecraft such as the Solar and Heliospheric Observatory (SOHO) \citep{domingo1995soho},  the Wind spacecraft \citep{ogilvie1995swe}, and the Parker Solar Probe \citep{fox2016solar,raouafi2022journey} have gathered detailed information about the solar wind's properties. However, these measurements are limited to specific points in space, and spacecraft are restricted in their proximity to the Sun, making it challenging to study the solar wind near the solar corona where it originates. To overcome these limitations and achieve a broader understanding of the solar wind's behaviour across the heliosphere, remote sensing techniques, such as interplanetary scintillation (IPS), are employed.\\

IPS describes the rapid fluctuations in the amplitude and phase of radio waves as they are scattered by the turbulent solar plasma \citep{hewish1964interplanetary}. This scattering occurs due to variations in the refractive index of the solar wind, influenced by plasma density. The solar wind, composed mainly of electrons and protons, is a stream of charged particles emitted by the Sun that fills the heliosphere. It is characterised by turbulent irregularities, leading to propagation effects including amplitude scintillation, phase scintillation, and spectral broadening \citep{coles1989propagation}. Studying these scintillations provides information along the entire path of the radio signal, offering a unique perspective on the solar wind's velocity, density fluctuations, and turbulence characteristics over a broad range of spatial scales \citep{hewish1964interplanetary,coles1978interplanetary}.\\

Previous studies have used the telemetry signals from spacecraft missions to study the solar wind. These spacecraft transmit radio signals that are affected by the solar wind, and the resulting phase and amplitude scintillations are analyzed to extract valuable information about the solar wind's properties. For example, observations of the European Space Agency's Venus Express mission by radio telescopes provided significant data on the solar wind by analyzing the phase scintillation of its X-band telemetry signal \citep{calves2014observations}, while the Mars Express mission extended these observations to different solar distances and phases of the solar cycle \citep{kummamuru2023monitoring}. Despite these valuable contributions, spacecraft observations are limited by their fixed trajectories and inability to achieve continuous and extensive spatial coverage.\\

Recent advancements in radio astronomy, particularly with wide-field instruments such as the Murchison Widefield Array \citep{morgan2018interplanetary}, ASKAP \citep{chhetri2023first}, and LOFAR \citep{van2013lofar}, have improved the study of IPS by offering simultaneous observations across multiple lines of sight with wide bandwidth at low frequencies. However, these arrays have limited coverage at higher frequencies. Very Long Baseline Interferometry (VLBI) can be used to estimate phase scintillation, as pointed out by \citet{cronyn1972interferometer}. Additionally, VLBI provides broader frequency coverage and enables observations closer to the Sun. VLBI has substantially enhanced our understanding of solar and interplanetary phenomena by observing natural radio sources, such as quasars, for IPS studies \citep[e.g.,][]{coles1989propagation, spangler1995radio, spangler2002very, spangler2003small}. VLBI potentially provides continuous monitoring and extensive spatial coverage of the solar wind's properties at a broad range of observing frequencies and can effectively estimate phase scintillation, providing critical insights into solar wind structure \citep{coles1989propagation}.\\

Geodetic VLBI sessions typically involve 7-12 globally distributed radio telescopes and observe hundreds of quasars \citep{nothnagel2017international}. The line of sight to many of the observed radio sources passes through the inner heliosphere at various positions around the Sun. This widespread spatial coverage allows for the capture of solar wind from multiple directions and lines of sight in a short period of time, offering a comprehensive understanding of its structure and dynamics. Consequently, the utilisation of geodetic VLBI networks enables the detection of anisotropies and asymmetries in the solar wind, such as those caused by coronal mass ejections (CMEs) \citep{spangler1995radio, molera2017analysis}. Additionally, it facilitates the reconstruction of three-dimensional structures in the solar wind \citep{tokumaru2013three}. \\

This study leverages the advanced capabilities of broadband VLBI to investigate the IPS phenomenon, focusing on phase scintillation analysis of observations to the radio sources 0003-066, 2126-158, and 2227-088, located near the Sun at varying solar elongations. Utilising the AuScope VLBI array, we aim to study solar wind turbulence through a detailed analysis of the phase scintillation index across multiple frequency channels. The observations are validated against the Kolmogorov turbulence model to capture the detailed characteristics of the solar wind, contributing to a broader understanding of heliophysics and the interplanetary medium. These objectives are driven by advancements in radio telescope technology and the desire to exploit significant investments in radio astronomy infrastructure. Integrating space weather observations into nominal geodetic VLBI sessions offers an opportunity to make space weather monitoring operational and routine, providing valuable data for both scientific research and space weather predictions.\\

This paper is structured as follows. Section \ref{sec2} outlines the methodology employed in this study. Section \ref{sec3} describes the observation setup and data processing, including the configuration of the AuScope VLBI array and the procedures for handling the data collected across multiple frequencies. In Section \ref{sec4}, we present the results and analysis, highlighting the derived phase scintillation indices. Finally, Section \ref{sec5} provides a discussion and conclusion, summarizing the main findings, exploring their significance in the context of solar and space weather phenomena, and suggesting potential directions for future research.\\

\section{Methodology}\label{sec2}

\begin{figure}[htbp!]
\begin{center}
\includegraphics[width=1\columnwidth]{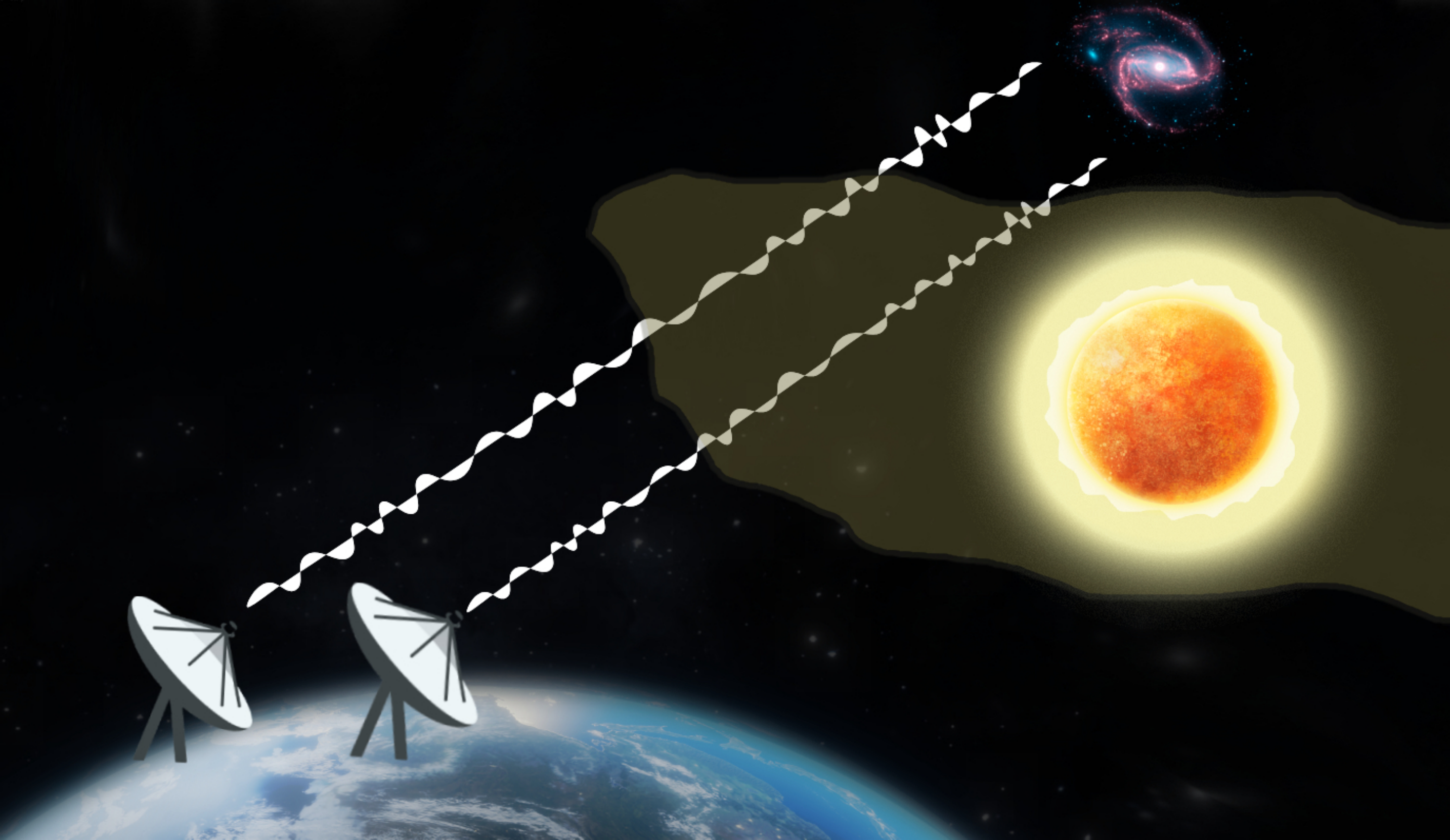}
\caption{Schematic of the observation principle for IPS using two radio telescopes, illustrating how radio waves from a natural radio source are affected by the solar plasma before reaching Earth.}
\label{FIG:sketch}
\end{center}
\end{figure}

Measuring IPS involves observing the scattered radio waves from distant compact radio sources, such as quasars, as they travel through the solar plasma. Figure \ref{FIG:sketch} illustrates how two radio telescopes capture these signals, which are influenced by solar plasma, and correlate them to generate visibility data composed of amplitude and phase measurements. By analyzing the phase fluctuations, information about solar wind properties can be obtained.\\

The geometry of the observation, including the baseline orientation and the relative position of the source with respect to the Sun, plays a critical role in determining the observed scintillation. Analyzing these patterns across varying baselines and frequencies allows for the separation of solar wind effects from other potential sources of scintillation, providing deeper insights into the properties of the intervening medium \citep{coles1978interplanetary}.\\ 

The phase of a radio wave that passes through the inner heliosphere directly correlates with fluctuations in the integrated electron content along the line of sight \citep{coles1989propagation}. In the context of VLBI, these phase fluctuations are measured between pairs of antennas (baseline) and are referred to as interferometric phase scintillation. The phase structure function is a fundamental tool for investigating turbulence in the interplanetary medium, particularly in characterizing the statistical properties of electron density fluctuations. \citet{spangler1995radio} demonstrated that the phase structure function can be directly computed from interferometer phase data using the phase scintillation index. In this study, we calculate the phase scintillation index as defined by \citet{duev2012spacecraft} and refined by \citet{calves2014observations} to provide a robust measure of the interferometric phase variation. Moreover, this study extends beyond the traditional single measurement of the power spectrum per band, instead utilising the large bandwidth of 3 to 13 GHz to generate 32 measurements across this frequency range, capturing a broad spectrum of IPS effects.\\

Phase scintillation refers to the rapid, random changes in the phase of a radio signal caused by irregularities in ionized plasma density. These phase scintillations, captured in interferometer phase data, provide important insights into the turbulence and density variations within the solar wind. The phase scintillation index ($\sigma_{\phi_{Sc}}$) is quantified as the variance of phase scintillation ($\phi_{Sc}$) derived from the filtered power spectrum (\(S_\phi(f)\)), expressed as:\\

\begin{equation}
    \phi_{Sc} = \left[ \int_{f_0}^{f_{max}} D(f_0) \cdot \left(\frac{f}{f_0}\right)^{m} df \right]^{1/2},
    \label{EQ:ips}
\end{equation}
\\

where \( D(f_0) \) denotes the spectral power density, \( f_0 \) is the low-frequency cutoff determined by the scan length, \( f_{max} \) is the high-frequency cutoff dictated by system noise, and $m$ is the slope of the spectral power density \citep{calves2014observations}. The phase scintillation measurements provide information on the plasma behaviour across all observed sky frequencies.\\

To further characterise the behaviour of the phase scintillation index across observing frequency, a power law model is employed. To minimize fitting errors, the frequencies were normalized by the reference frequency (8 GHz), which lies in the middle of the frequency range, ensuring that the domain values are distributed symmetrically around zero.\\

\begin{equation}
     \sigma_{\phi_{Sc}}(\nu) = A \cdot \left(\frac{8~GHz}{\nu}\right) ^{-\alpha},
    \label{EQ:slope}
\end{equation}
\\

where $\nu$ is the observing frequency, and $A$ is the phase scintillation index at the reference frequency, providing an averaged value that smooths out noise and anomalies.  The power law index ($\alpha$) describes how the phase scintillation index changes with observing frequency. This approach is more robust than relying on a single value at the reference observing frequency. Therefore, $A$ is used to represent the IPS behaviour against the elongation angle, and $\alpha$ illustrates how the phase scintillation index changes with increasing solar elongation.\\

By using 32 channels across a broad frequency range, the phase scintillation index captures a wide range of scattering strengths, accommodating both high scattering at lower frequencies (where phase changes may be rapid) and weaker scattering at higher frequencies (where phase changes may be subtle). This capability results in more robust data with higher signal-to-noise ratios. These multi-frequency measurements thus provide a more comprehensive understanding of IPS behaviour, allowing more accurate model fitting across frequencies, and enabling analyses that cover a broader spectrum of solar wind turbulence characteristics.\\


\section{Observation setup and data processing}\label{sec3}


The AuScope VLBI Array was constructed as part of the National Cooperative Research Infrastructure Strategy (NCRIS), funded by the Australian Department of Innovation, Industry, Science, and Research \citep{NCRIS}. It consists of three 12-meter radio telescopes located in Hobart-Tasmania (Hb), Katherine-Northern Territory (Ke), and Yarragadee-Western Australia (Yg) \citep{lovell2013}. The VLBI array is a component of AuScope, a comprehensive infrastructure framework for studying Earth's geology, chemistry, physics, and geography \citep{auscope}, and operated by the University of Tasmania. The array was designed to be a fast-slewing, high-precision system and it is a key contributor to the International VLBI Service for Geodesy and Astrometry (IVS) \citep{nothnagel2017international}, which coordinates the global VLBI observations. \\

\begin{figure}[htbp!]
\begin{center}
\includegraphics[width=1\columnwidth]{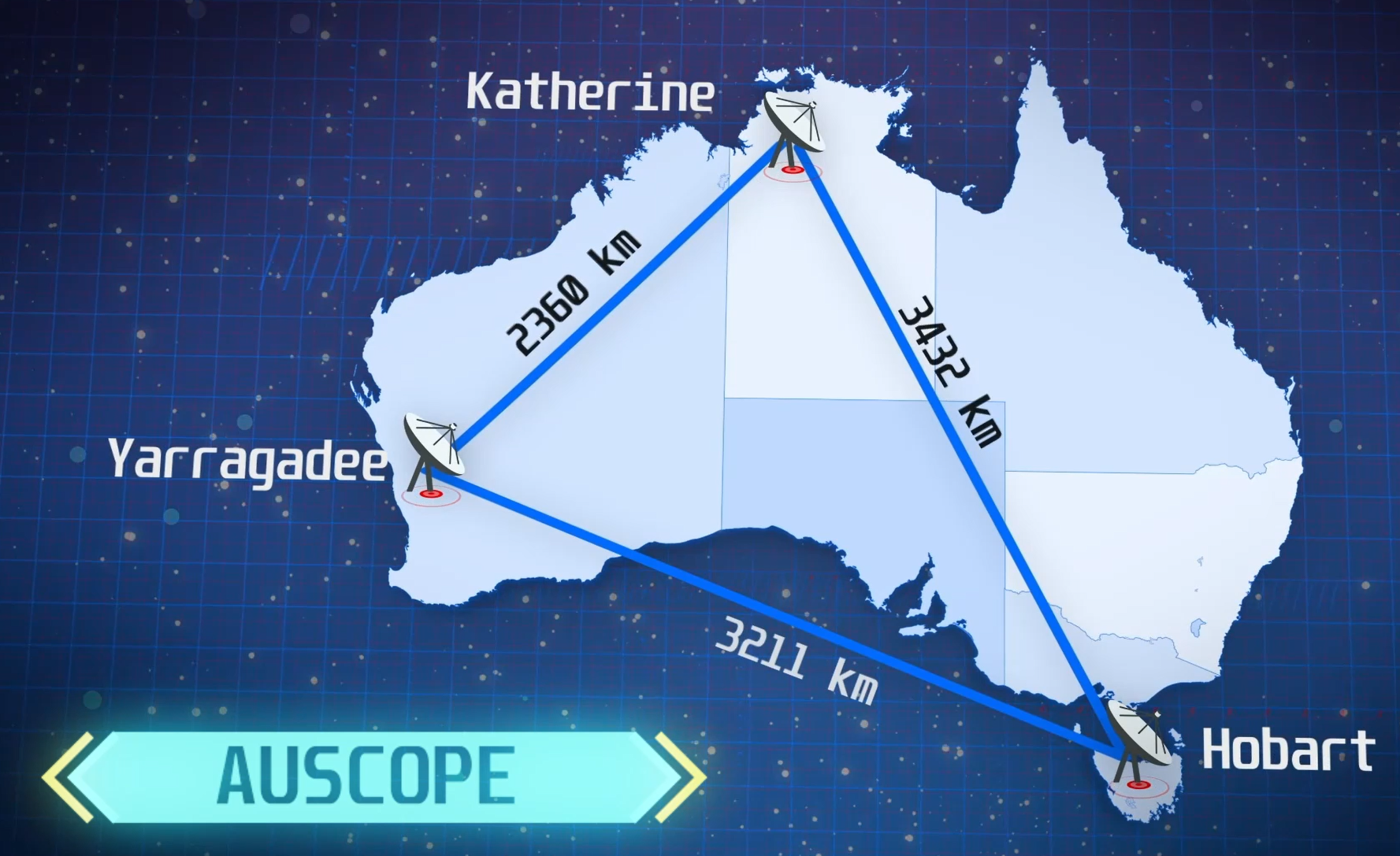}
\caption{The AuScope VLBI array, consisting of Hobart 12m (Hb) in Tasmania, Katherine 12m (Ke) in the Northern Territory, and Yarragadee 12m (Yg) in Western Australia.}
\label{FIG:auscope}
\end{center}
\end{figure}

\subsection{Details on the Broadband system - Technical overview}

The array utilises a pedestal-mounted telescope design with a 12-meter Cassegrain dish and a 1.8-meter sub-reflector with a slew rate up to $5^\circ$/s in azimuth and $1.25^\circ$/s in elevation \citep{lovell2013}. Initially, however, both the front-end and back-end were designed in the legacy S/X observing mode, i.e., two discrete bands at 2.4 GHz and 8.4 GHz. An upgrade to the receiver and back-end systems to enable broadband observations was completed in 2024.\\


The broadband system was developed as a new technical realization of the geodetic VLBI hardware,  operating across multiple GHz of bandwidth. The front-end includes a wideband feed and low-noise amplifiers, designed to receive a broad range of frequencies. The front-end was designed by Callisto \footnote{\url{https://www.callisto-space.com/en/}} with a QRFH feed that produces two linear polarizations within the frequency range 2-14 GHz. It also includes cooling systems; Stirling cycle cooling, to reduce thermal noise and improve the performance of the receiver \citep{akgiray2012circular}. A High Pass Filter (HPF) filters the signal, allowing only frequencies above 3 GHz to pass to reduce the Radio Frequency Interference (RFI) at lower frequencies.\\

\begin{figure}[htbp!]
\begin{center}
  \includegraphics[width=1\textwidth,trim={1cm 3cm 1cm 3cm},clip]{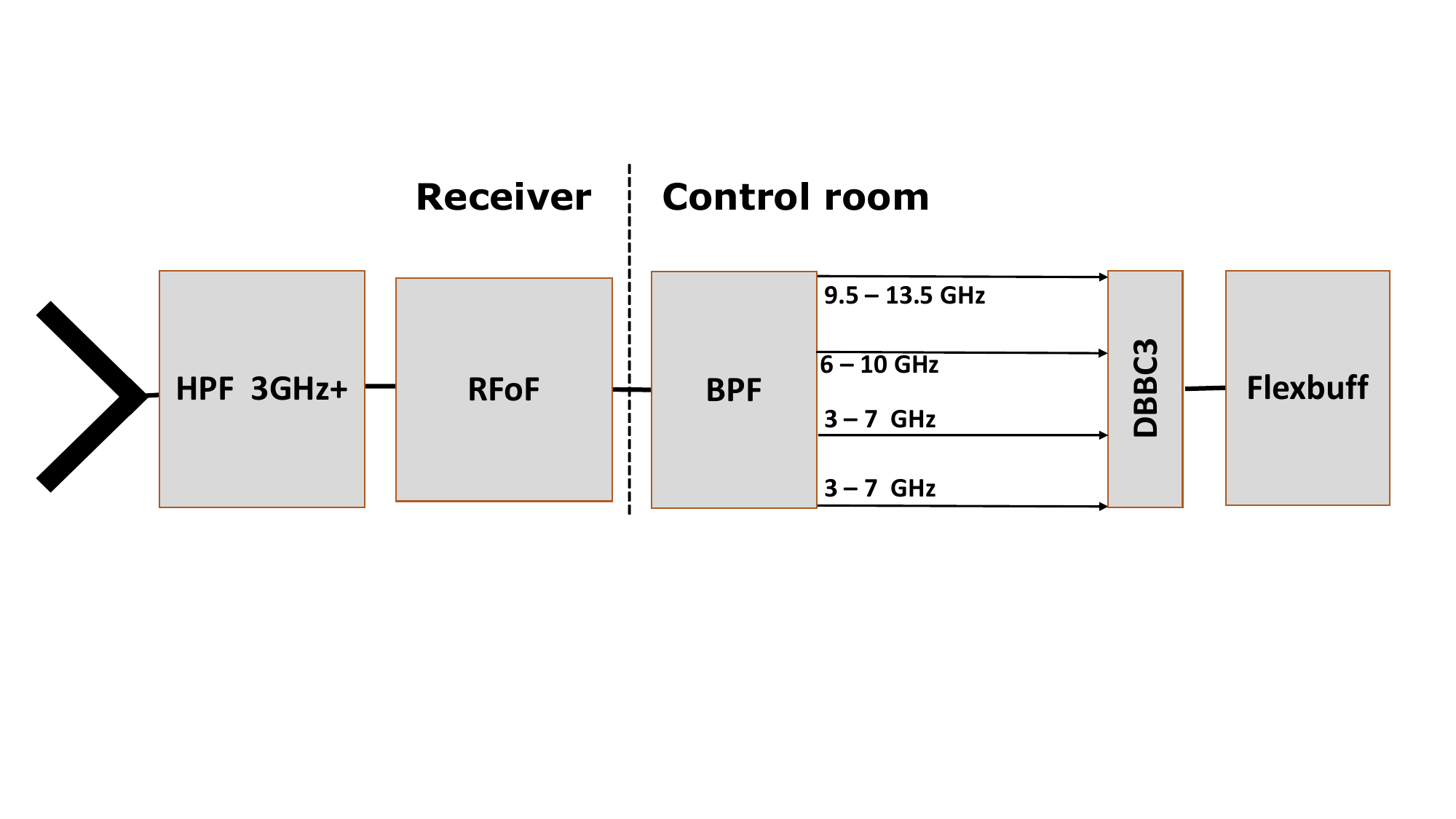}
  \caption{The signal chain for the broadband system; the signal received by the QRFH feed, filtered with HPF, and conveyed signal larger than 3 GHz to the control room via RFoF for down-conversion and sampling at DBBC3 before being recorded by Flexbuff.}
  \label{FIG:signal}
\end{center}
\end{figure}

The signal is transmitted via Radio Frequency over Fibre (RFoF) to the control room, where it is bandpass filtered into four distinct bands. Each band has an input bandwidth of 4 GHz, covering frequency ranges of 3-7 GHz, 6-10 GHz, and 9.5-13.5 GHz, with the first band being duplicated. These bands are simultaneously down-converted and sampled by the DBBC3 system \citep{tuccari2018dbbc3}. The data are then recorded using jive5ab\footnote{\url{https://github.com/jive-vlbi/}} onto a Flexbuff machine in VDIF format (see Figure \ref{FIG:signal}). The baseband channels can be flexibly selected within the band, offering bandwidths from 8 to 64 MHz and allowing up to 16 channels per band.\\

\subsection{Observation setup}
The observation mode employed 32 channels (8 channels per band), each with a bandwidth of 32 MHz and two polarizations, resulting in a recording rate of 8 Gbps per antenna. This setup mimics the Geodetic VLBI observing mode used in the broadband configuration to ensure IPS studies can use standard geodetic observation data.  A selection of geodetic sources with flux densities exceeding 0.5 Jy and located within $12^\circ$ of solar elongation were selected. Observations were carried out over multiple epochs, with sources of interest (hereafter referred to as primary sources) being observed for a duration of 480 seconds multiple times each session, with a session duration of 1-hour. See Table \ref{tab:sched} for details.\\

In addition to the primary sources, calibration sources located far from the Sun were observed for 180 seconds. These calibration sources were used in the manual phase calibration process as discussed in subsection \ref{subsec33}. Furthermore, these calibration sources ensure the accuracy and reliability of the data by providing a benchmark for comparison against the primary sources.  The scheduling strategy involved alternating between calibration sources and primary sources.\\ 

\begin{table*}[htbp!]
\begin{tabular}{ccccc}
\toprule
 \headrow \textbf{Epoch}         & \textbf{Session}       & \textbf{Source}    & \textbf{\begin{tabular}[c]{@{}c@{}}Solar elongation\\  {[}degree{]}\end{tabular}} & \textbf{Comment}    \\ \midrule
\multirow{2}{*} {\raisebox{0.4ex}{17-02-2024}} & \multirow{2}{*}{\raisebox{0.4ex}{Q0217}} & 2126-158           & 7.8                                                                               & Primary source      \\ \cmidrule(l){3-5} 
                       &                        & 2255-282, 1921-293 &                                                                                   & Calibration sources \\ \midrule
\multirow{2}{*}{\raisebox{0.4ex}{18-02-2024}} & \multirow{2}{*}{\raisebox{0.4ex}{Q0218}} & 2227-088           & 7.2                                                                               & Primary source      \\ \cmidrule(l){3-5} 
                       &                        & 2255-282, 1921-293 &                                                                                   & Calibration sources \\ \midrule
\multirow{2}{*}{\raisebox{0.4ex}{10-03-2024}} & \multirow{2}{*}{\raisebox{0.4ex}{Q0310}} & 0003-066           & 11.3                                                                              & Primary source      \\ \cmidrule(l){3-5} 
                       &                        & 2255-282, 1921-293 &                                                                                   & Calibration sources \\ \midrule
\multirow{2}{*}{\raisebox{0.4ex}{11-03-2024}} & \multirow{2}{*}{\raisebox{0.4ex}{Q0311}} & 0003-066           & 10.5                                                                              & Primary source      \\ \cmidrule(l){3-5} 
                       &                        & 2255-282, 1921-293 &                                                                                   & Calibration sources \\ \midrule
\multirow{2}{*}{\raisebox{0.4ex}{20-03-2024}} & \multirow{2}{*}{\raisebox{0.4ex}{Q0320}} & 0003-066           & 6.5                                                                               & Primary source      \\ \cmidrule(l){3-5} 
                       &                        & 2255-282, 1921-293 &                                                                                   & Calibration sources \\ \midrule
\multirow{2}{*}{\raisebox{0.4ex}{23-03-2024}} & \multirow{2}{*}{\raisebox{0.4ex}{Q0323}} & 0003-066           & 7.4                                                                               & Primary source      \\ \cmidrule(l){3-5} 
                       &                        & 2255-282, 1921-293 &                                                                                   & Calibration sources \\ \midrule
\multirow{2}{*}{\raisebox{0.4ex}{25-03-2024}} & \multirow{2}{*}{\raisebox{0.4ex}{Q0325}} & 0003-066           & 8.5                                                                               & Primary source      \\ \cmidrule(l){3-5} 
                       &                        & 2255-282, 1921-293 &                                                                                   & Calibration sources \\ \bottomrule
\end{tabular}
\caption{Summary of Observations schedule}
\label{tab:sched}
\end{table*}

\begin{figure*}[htbp!]

\begin{center}
  \includegraphics[width=0.8\textwidth,trim={2.5cm 0cm 2.5cm 0cm},clip]{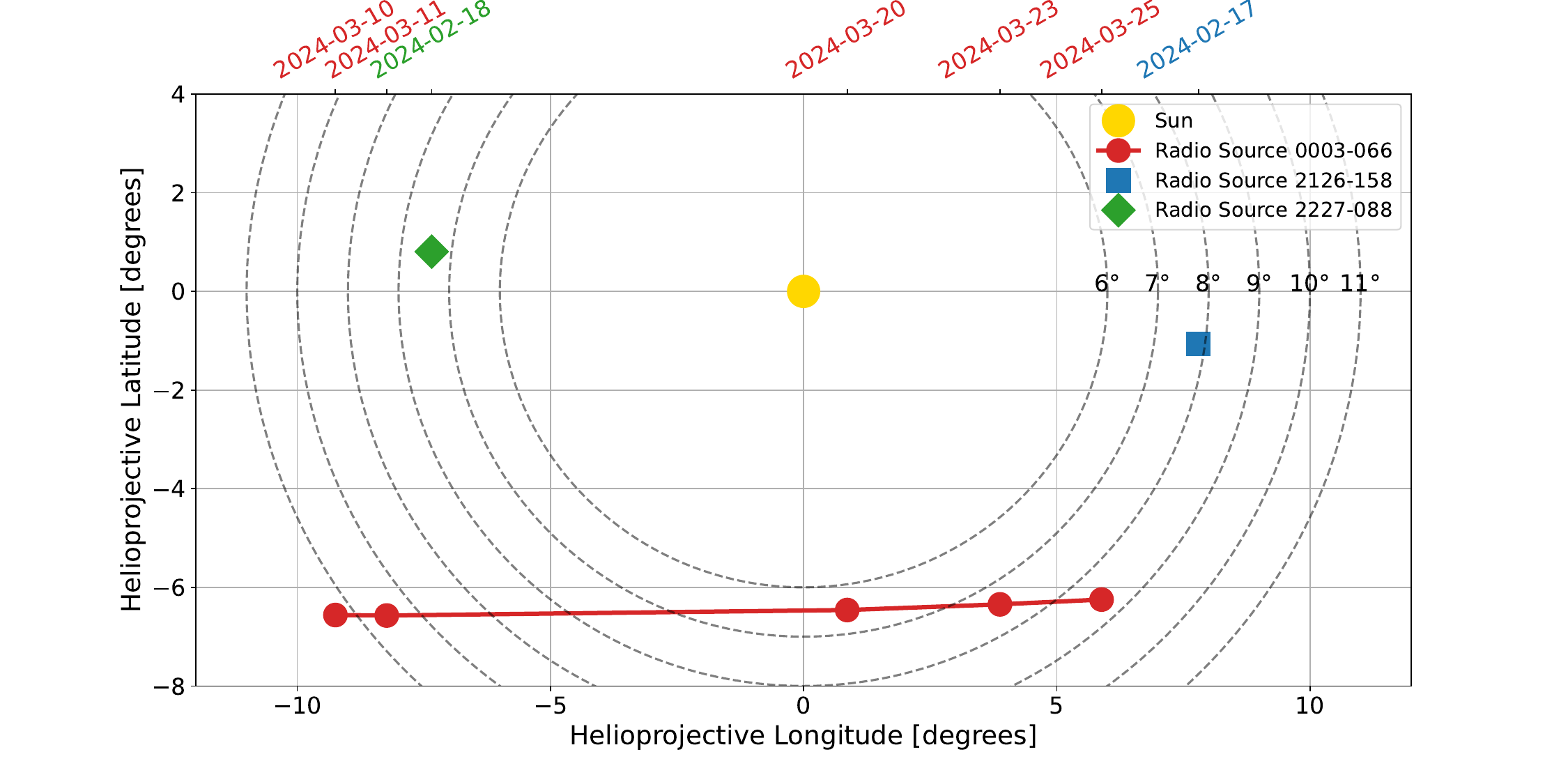}
  \caption{Helioprojective coordinates of the radio sources 0003-066, 2126-158, and 2227-088 relative to the Sun on specified observation dates. The Sun is represented by the gold circle, and concentric black dashed circles indicate elongation angles from the Sun. The red circles connected by lines show the positions of 0003-066 over the observation period, while the square and diamond represent the positions of 2126-158 and 2227-088.}
  \label{FIG:SourcePath}
\end{center}
\end{figure*}

Figure \ref{FIG:SourcePath} depicts the helioprojective coordinates of the radio source 0003-066 on five different observation dates (10-03-2024, 11-03-2024, 20-03-2024, 23-03-2024, and 25-03-2024) relative to the Sun, as well as the sources 2126-158 and 2227-088 on 17 February 2024 and 18 February 2024, respectively. The Sun is shown as the gold circle at the centre of the plot. The red circles connected by lines indicate the positions of the source 0003-066 on the respective dates, while the sources 2126-158 and 2227-088 are represented by a square and a diamond, respectively. Concentric black dashed circles indicate elongation angles of $6^\circ$, $7^\circ$, $8^\circ$, $9^\circ$, $10^\circ$, and $11^\circ$ from the Sun. The secondary x-axis shows the observation dates corresponding to each plotted position. This visualization aids in understanding the relative positions and motions of the radio sources with respect to the Sun over the specified period.\\

\begin{table}[htb!]
\begin{threeparttable}
\caption{Comparison of flux densities for the observed primary and calibration sources at X-band, as provided by the SKED catalog and internally estimated. }
\begin{tabular}{cccc}
\toprule
\headrow \textbf{Source} & \textbf{\begin{tabular}[c]{@{}c@{}}Nominal flux  \\ (SKED){[}Jy{]} \end{tabular}} & \textbf{\begin{tabular}[c]{@{}c@{}}Estimated flux\\ {[}Jy{]}\end{tabular}} & \textbf{Comment}   \\ \midrule
2126-158        & 2.2                                                                              & 0.55                                                                       & Primary source     \\
\midrule
2227-088        & 1.19                                                                             & 0.5                                                                        & Primary source     \\
\midrule
0003-066        & 4.97                                                                             & 1.7                                                                        & Primary source     \\
\midrule
1921-293        & 10                                                                               & 7.4                                                                        & Calibration source \\
\midrule
2255-282        & 3.51                                                                             & 1.8                                                                        & Calibration source \\ \bottomrule
\end{tabular}
\label{tab:flux}
\end{threeparttable}

\end{table}

According to the SKED scheduling software's \texttt{flux.cat} \citep{gipson2018sked}, the X-band flux densities for the primary sources 0003-066, 2126-158, and 2227-088 are 5 Jy, 2.2 Jy, and 1.2 Jy, respectively. However, an internal session conducted on 30-09-2023 over the Hb-Ke baseline estimated their apparent flux densities at X band to be 1.5 Jy, 0.6 Jy, and 0.5 Jy\footnote{Such discrepancies are not unusual in VLBI due to source variability, and differences in baseline lengths and orientation.}, respectively, with a System Equivalent Flux Density (SEFD) of 5000 Jy for this baseline (see Table \ref{tab:flux}). Consequently, we adopted the flux values obtained from the internal session.\\

\begin{figure*}[htbp!]

\begin{center}
  \includegraphics[width=1\textwidth,trim={0cm 0cm 0cm 0cm},clip]{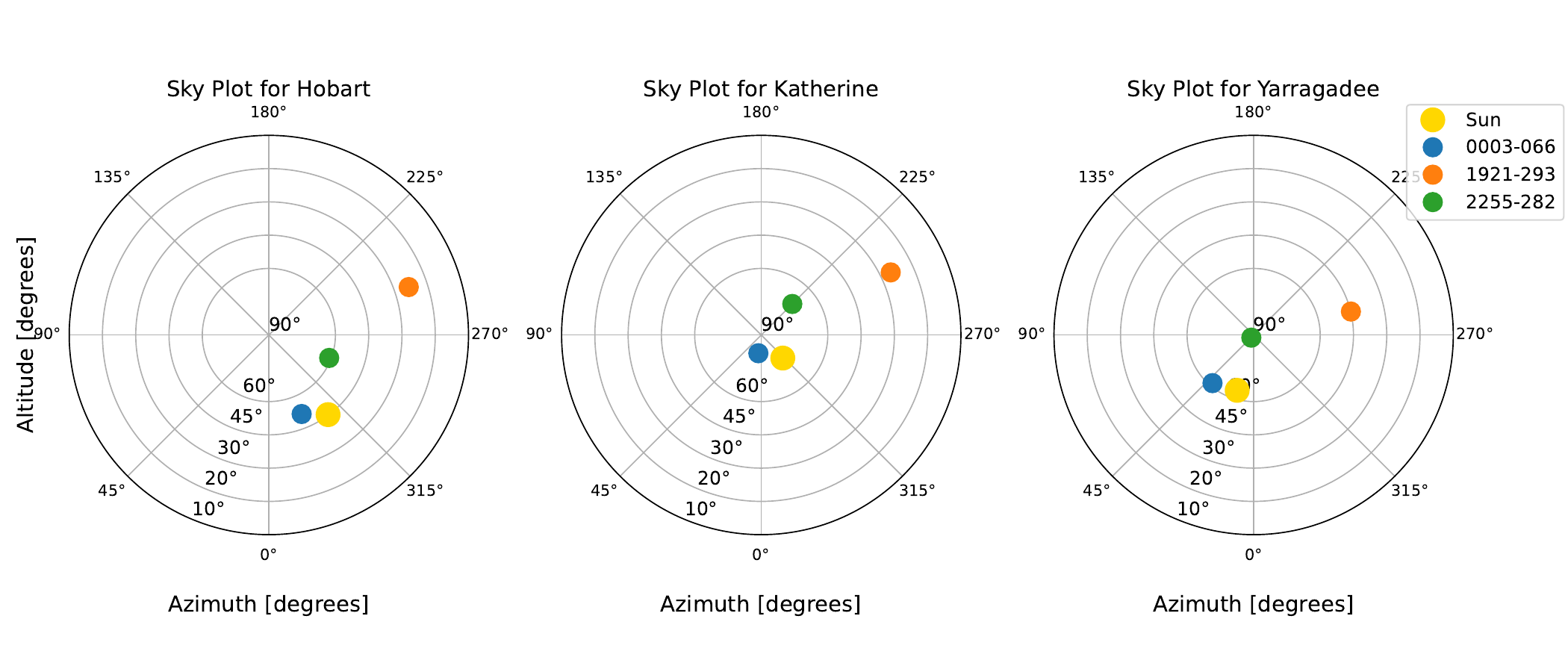}
  \caption{Sky plots of three radio sources (primary source and calibration sources) and the Sun from three stations during session Q0310.}
  \label{FIG:skyplot}
\end{center}
\end{figure*}

The scheduling also accounted for a cutoff elevation angle, ensuring primary sources were observed at elevations greater than 20 degrees and calibration sources at elevations greater than 10 degrees to maintain a low SEFD for the stations. Polar sky plots illustrating the positions of three radio sources (0003-066, 1921-293, and 2255-282) and the Sun, as observed from three different stations on 23-03-2024 at 02:00 UT, are shown in Figure \ref{FIG:skyplot}. Each subplot presents the azimuth and altitude of the sources and the Sun, with the Sun depicted as a gold circle with a black edge for enhanced visibility. The red, blue, and green markers indicate the positions of the radio sources 0003-066, 1921-293, and 2255-282, respectively. To facilitate the comparison of source positions across different observatories, the plots share the same y-axis for altitude. The varying distances between the Sun and the sources in each subplot illustrate the parallax effect due to the geographic locations of the observatories. \\

\subsection{Correlation and fringe fitting}\label{subsec33}

The data were shipped from Ke and e-transferred from Yg to Hb to be correlated at UTAS using DiFX \citep{deller2011difx}.  The correlator outputs cross-correlation coefficients for each scan, which are post-processed to estimate fringe amplitude, phase, group delay, phase delay rate, and differential Total Electron Content (dTEC) along the line of sight to a radio source. \\

Due to the Automatic Gain Controller (AGC) adjustment, the first 50 seconds of all scans following the initial scan were unstable and therefore discarded before correlating the data. The correlator simultaneously processes data from the four frequency bands, which involves reorganizing the data streams into files and integrating additional information such as station locations and Earth Orientation Parameters (EOP). Although the input parameters are accurate, uncertainties in UTC time tags require adjustments during the correlation process. The primary objective is to minimize post-correlation residual delays to below 100 ns by fine-tuning the station clocks based on initial correlation outcomes \citep{niell2018demonstration}. The final output consists of a series of complex correlation coefficients for each scan, covering various delays, frequency bands, and polarizations.\\

Post-processing of the correlator's output involves determining the group delay and phase delay rate values that maximize the delay/delay rate resolution function, using a maximum-likelihood approach \citep{rogers1970very}. This essential step, performed with the Haystack Observatory Post-processing System (HOPS) software \citep{HOPS}, is known as fringe fitting. Broadband fringe fitting requires multiple runs of the \texttt{fourfit} program under different configurations to correct channel phase offsets, synthesize bands, and combine polarizations, ultimately generating a set of calibrated broadband VGOS observables for each scan \citep{cappallo2017fourfit}. Calibration involves applying delay and phase corrections to each frequency channel to account for instrumental effects, a process referred to as phase calibration or instrumental delay calibration. Typically, calibration tones are injected at the front-end to align delays across bands, with adjustments made to ensure that single-band delays agree within 100 ns. However, at AuScope stations, the calibration tones are excessively strong at low frequencies, causing saturation effects. As a result, manual phase calibration (\texttt{pc\_mode manual}) was performed using a high signal-to-noise ratio (SNR) scan of a calibration source. Broadband group delay and phase are calculated by combining the complex polarization correlation coefficients across all frequency bands. Achieving coherent combinations requires correcting for delay and phase differences between polarizations. The output after the correlation and fringe fitting is interferometric visibility measurement as complex numbers characterised as amplitude and phase with 1-second averaging. In this paper, we studied the effect of the solar plasma using only the phase measurements.  \\

\subsection{Post-analysis}

The interferometric visibility data produced after fringe fitting spans both frequency (3 to 13 GHz) and baseline dimensions (Hb-Ke, Hb-Yg, Ke-Yg), allowing for analysis of IPS effects across these parameters. This dual-dimensional dataset examines how phase scintillation varies across frequency channels and baseline lengths, providing a detailed view of IPS behaviour and solar wind turbulence characteristics.\\

\begin{figure*}[htbp!]
\begin{center}
  \includegraphics[width=0.6\textwidth,trim={2cm 1cm 2cm 1cm},clip]{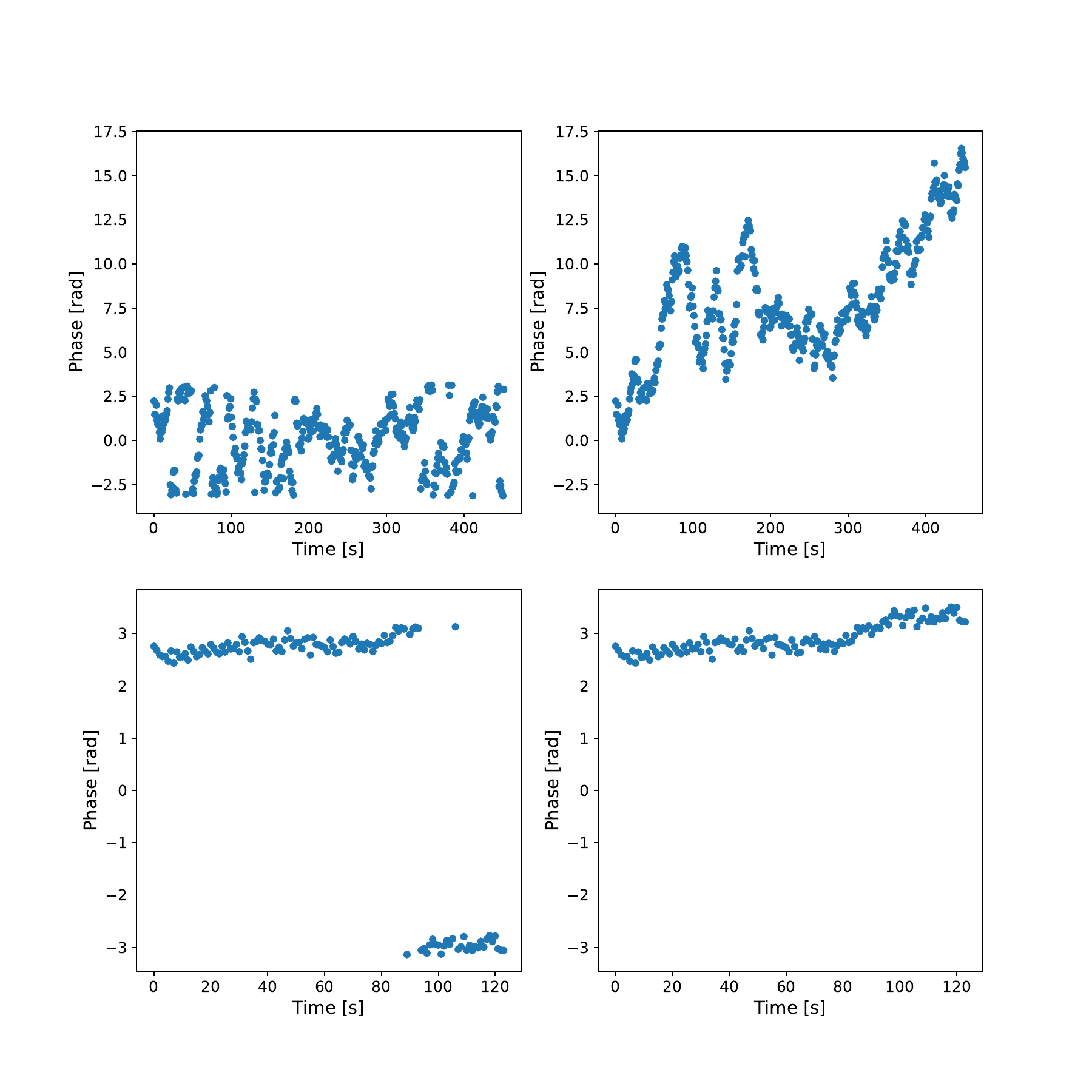}
  \caption{Phase time series for the primary source (top panels) and calibration source (bottom panels) at 5GHz. The left panels show the phase time series before unwrapping. The right panels show the phase time series after unwrapping.}
  \label{FIG:unwrap}
\end{center}
\end{figure*}

Once the fringe fitting is complete, the phase fluctuations in the 7-minute scan length were mainly due to IPS, radiometer noise, ionospheric scintillation, and tropospheric variations. The phase data clearly showed fluctuations caused by IPS, as illustrated in the left panels of Figure \ref{FIG:unwrap}. The top-left panel presents the phase of a primary source at 5 GHz, and the bottom-left panel presents the phase of a calibration source at the same frequency. Both scans were affected by the same radiometer noise but different ionospheric and tropospheric conditions, as the telescopes were pointing at different parts of the sky. The ionosphere and troposphere can cause phase fluctuations, and it is conceivable that tropospheric and ionospheric effects might be slightly different between primary and calibration sources. However, such fluctuations cannot explain fluctuations of the magnitude observed here, leaving IPS as the primary cause of the significant differences between the two panels. This is evidenced by the stark difference between the phase behaviour observed in the calibrator source, compared to the primary source.\\

The interferometric phase data is prone to phase discontinuities, or jumps, due to phase wrapping. This phenomenon occurs because the interferometric phase ranges between $\pi$ and $-\pi$, resulting in $2\pi$ discontinuities. The unwrapping process begins by examining the phase differences between consecutive data points. If a phase difference exceeds $\pi$, it is assumed that a $2\pi$ discontinuity has occurred. An appropriate multiple of $2\pi$ is then added or subtracted from the subsequent phase values to correct for this jump. This procedure is repeated iteratively across the entire phase time series, effectively smoothing out the phase discontinuities and reconstructing the continuous phase signal \citep{spangler1995radio}. The right panels in Figure \ref{FIG:unwrap} show the phase time series after unwrapping. The bottom-left panel displays the phase of the calibration source, where the phase jump due to the wrapping effect is clear after 90 seconds. The phase jumps due to the wrapping effect are corrected, resulting in a continuous phase signal as seen in the right panels.\\

\begin{figure*}[htbp!]
\begin{center}
  \includegraphics[width=0.8\textwidth,trim={1cm 0cm 1cm 1cm},clip]{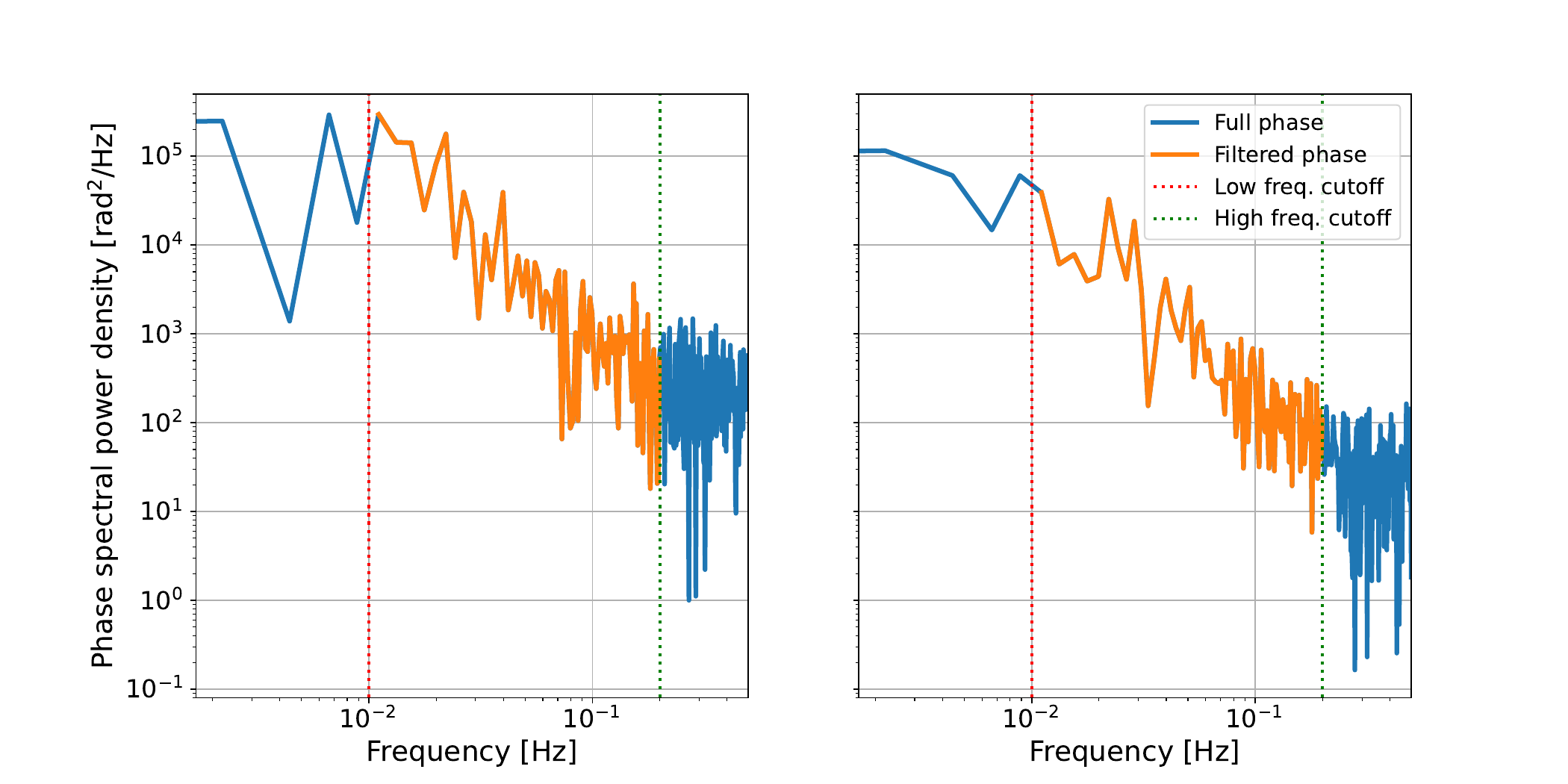}
  \includegraphics[width=0.8\textwidth,trim={1cm 0cm 1cm 1cm},clip]{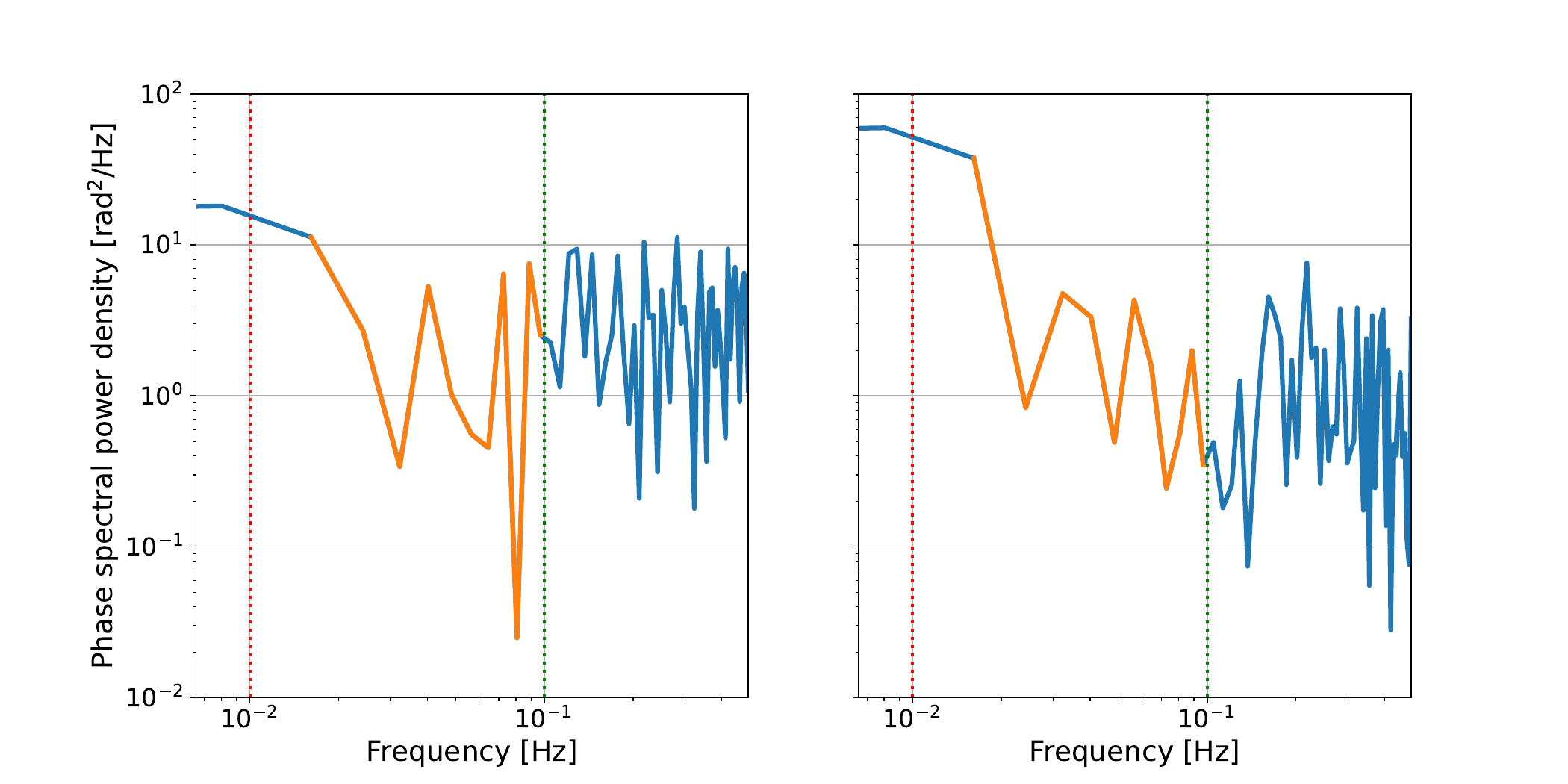}
  \caption{Power spectra of the phase residuals as a function of frequency at 3 GHz (left panel) and 5 GHz (right panel). The top panels show the power spectra of the primary source, while the bottom panels display the power spectra of the calibration source.}  The blue line represents the full-phase power spectrum. The dashed red and green lines indicate the low and high-frequency cutoffs, respectively. The orange highlighted region indicates the frequency range from 0.01 Hz to 0.2 Hz (scintillation region), where the IPS effect is significant. Frequencies above 0.2 Hz are dominated by system noise.
  \label{FIG:spectra}
\end{center}
\end{figure*}

To remove the influence of the long-timescale phenomena, a first-degree polynomial was fitted to the phase time series, yielding phase residuals. The phase scintillation index was subsequently derived from the phase residual power spectrum, with a Nyquist frequency of 0.5 Hz. The phase residual power spectra for all scans across different channels were thoroughly analyzed, revealing that the IPS effect spanned the frequency range from 0.01 to 0.2 Hz. This frequency range was applied to all scans across all channels to maintain the consistency of the results. Frequencies above 0.2 Hz are dominated by system noise. Figure \ref{FIG:spectra} illustrates the power spectra of the primary source (top panels) and calibration source (bottom panel). The filtered power spectrum is shown in orange, while the full-phase power spectrum is shown in blue. The dashed red and green lines indicate the low and high-frequency cutoffs, respectively. The left panels show the phase residual power spectrum at 3 GHz, while the right panels show it at 5 GHz. As expected, the power level of the primary source at 3 GHz is higher than at 5 GHz, consistent with the stronger impact of IPS at lower observing frequencies, as discussed in detail in Section \ref{sec4}. In contrast, the power level of the calibration source is significantly lower than that of the primary source due to the absence of the IPS impact. Additionally, there is no significant change in the power level of the calibration source between 3 GHz and 5 GHz. \\ 

Figure \ref{FIG:spectra} illustrates the power spectra of the primary source (top panels) and the calibration source (bottom panels). The filtered power spectrum is shown in orange, while the full-phase power spectrum is shown in blue. The dashed red and green lines represent the low- and high-frequency cutoffs, respectively. The left panels display the phase residual power spectrum at 3 GHz, while the right panels display it at 5 GHz. As expected, the power level of the primary source at 3 GHz is higher than at 5 GHz, consistent with the stronger impact of IPS at lower observing frequencies, as discussed in detail in Section \ref{sec4}. In contrast, the power level of the calibration source is significantly lower than that of the primary source due to the absence of IPS impact. Furthermore, there is no significant change in the power level of the calibration source between 3 GHz and 5 GHz.

\section{Results and Analysis}\label{sec4}

\begin{figure*}[htbp!]
\begin{center}
\includegraphics[width=0.8\columnwidth,trim={0.45cm 0cm 0.45cm 0cm},clip]{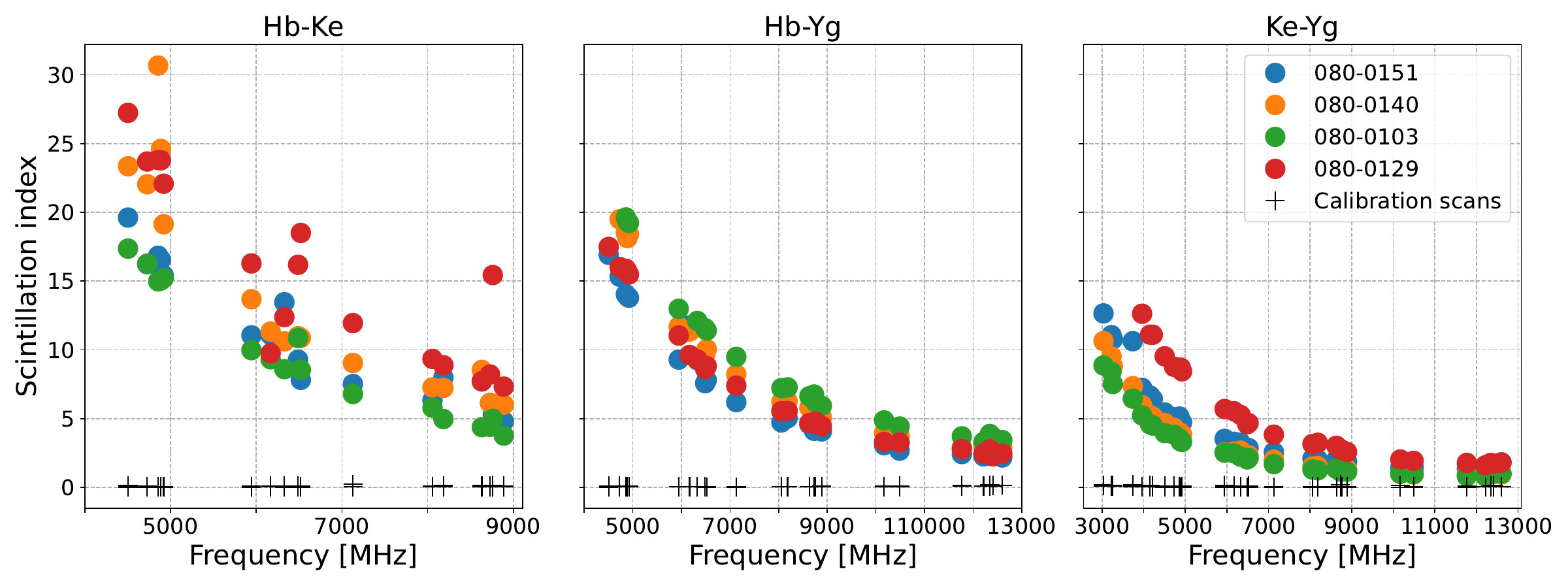}
\caption{The phase scintillation index for all three baselines (Hb-Ke, Hb-Yg, Ke-Yg) during session Q0320, plotted against frequency. Filled circles represent the phase scintillation index for the primary source, while the black $+$ markers represent calibration sources. Different colours correspond to different scans of the primary source during the session.}
\label{FIG:ips}
\end{center}
\end{figure*}

Using Equation \ref{EQ:ips}, the phase scintillation index was calculated for all channels across all scans and sessions. Figure \ref{FIG:ips} presents the phase scintillation index for the three baselines, Hb-Ke, Hb-Yg, and Ke-Yg, during session Q0320 for the primary source, plotted against the channels. The phase scintillation index shows a decreasing trend with increasing frequency, which aligns with theoretical models. Outliers, particularly for the Hb-Ke baseline beyond 9 GHz, mainly due to system noise, were removed from the plot to improve clarity. Hb station experienced a lack of one polarization from 3-4.5 GHz during this session, leading to the exclusion of these data points from the Hb baselines to ensure consistency in the analysis. The differences between scans in Figure \ref{FIG:ips} do not indicate systematic offsets.  These variations are more likely attributable to system noise than to factors like baseline rotation. Additionally, the short session duration minimizes any potential influence of baseline rotation.\\

Data from different scans over the 1-hour session show slight magnitude variations but follow the same trend, demonstrating measurement reliability. However, the Hb-Ke baseline showed high uncertainty and numerous outliers, primarily due to its poor SEFD, which is lower than the other baselines. The Hb station's low-sensitivity significantly affects Hb-Ke baseline performance, while the Yg station's high-sensitivity enhanced the performance of the Hb-Yg and Ke-Yg baselines. This highlights the crucial role of station sensitivity in acquiring precise IPS measurements.\\

In addition to the primary source, the phase scintillation index for the calibration sources is also plotted using the + markers in Figure \ref{FIG:ips}. The calibration sources were processed in the same manner as the primary source and exhibit a very low scintillation index, primarily due to radiometer noise, with values consistently below 0.5. This distinct behaviour highlights the lower scintillation for calibration sources compared to the primary source. At higher frequencies, the radiometer noise increases, and as the phase scintillation index decreases, the radiometer noise becomes more dominant, flattening the curve of phase scintillation versus frequency. \\

To understand the relationship between phase scintillation index and frequency, the power law model introduced in Equation \ref{EQ:slope} was applied. Frequencies were normalized by the reference frequency of 8 GHz to minimize fitting errors and provide a balanced distribution. This model produces an averaged phase scintillation index $A$ at 8 GHz, which smooths out noise and anomalies, and a power law index $\alpha$, capturing the frequency dependence of the scintillation index.\\

\begin{figure}[htbp!]
\begin{center}
\includegraphics[width=1\columnwidth,trim={1cm 0cm 1cm 0cm},clip]{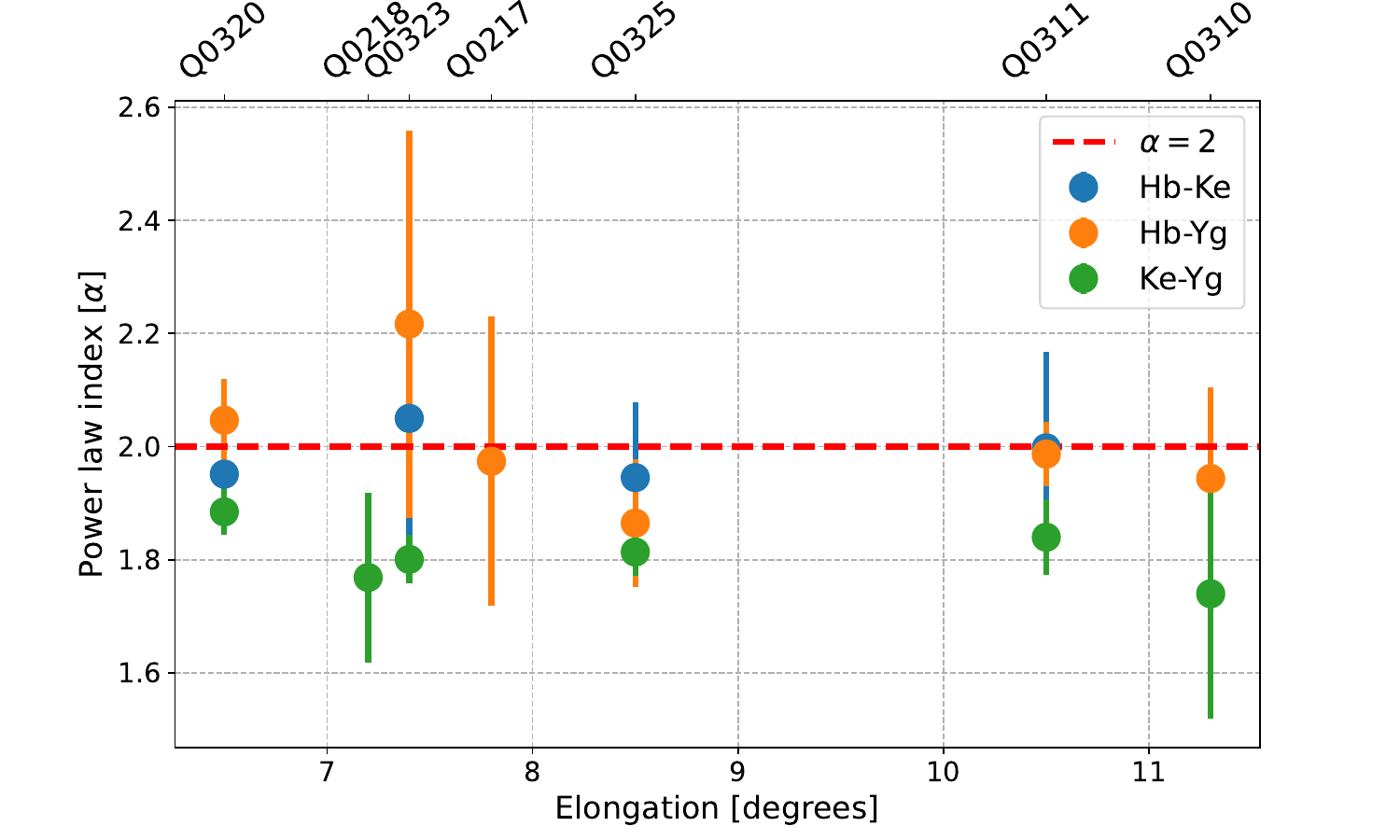}
\caption{Power law index ($\alpha$) as a function of elongation angle for the three baselines (Hb-Ke, Hb-Yg, Ke-Yg). Data points are shown with error bars, where circles represent the mean power law index across all scans at each elongation angle, and the error bars indicate the standard deviation. The red dashed line represents the expected quadratic power law.}
\label{FIG:quad_elong}
\end{center}
\end{figure}

The phase scintillation index is expected to follow a quadratic power law with frequency. Figure \ref{FIG:quad_elong} presents the power law index ($\alpha$) for all scans. The values are generally close to the expected quadratic index. This supports the validity of using the broadband system in capturing IPS data across a wide range of frequencies. However, the Ke-Yg baseline consistently shows lower power law indices compared to other baselines. This can be attributed to system noise, which becomes more prominent at higher frequencies for Ke-Yg due to its lower IPS values. As shown in Figure \ref{FIG:quad_elong}, the phase scintillation index for Ke-Yg approaches the level of system noise at high frequencies, leading to a steeper deviation from the expected power law behaviour.  \\

\subsection{Phase scintillation index as a function of elongation}

\begin{figure}[htbp!]
\begin{center}
\includegraphics[width=1\columnwidth,trim={1.5cm 0cm 1.5cm 0cm},clip]{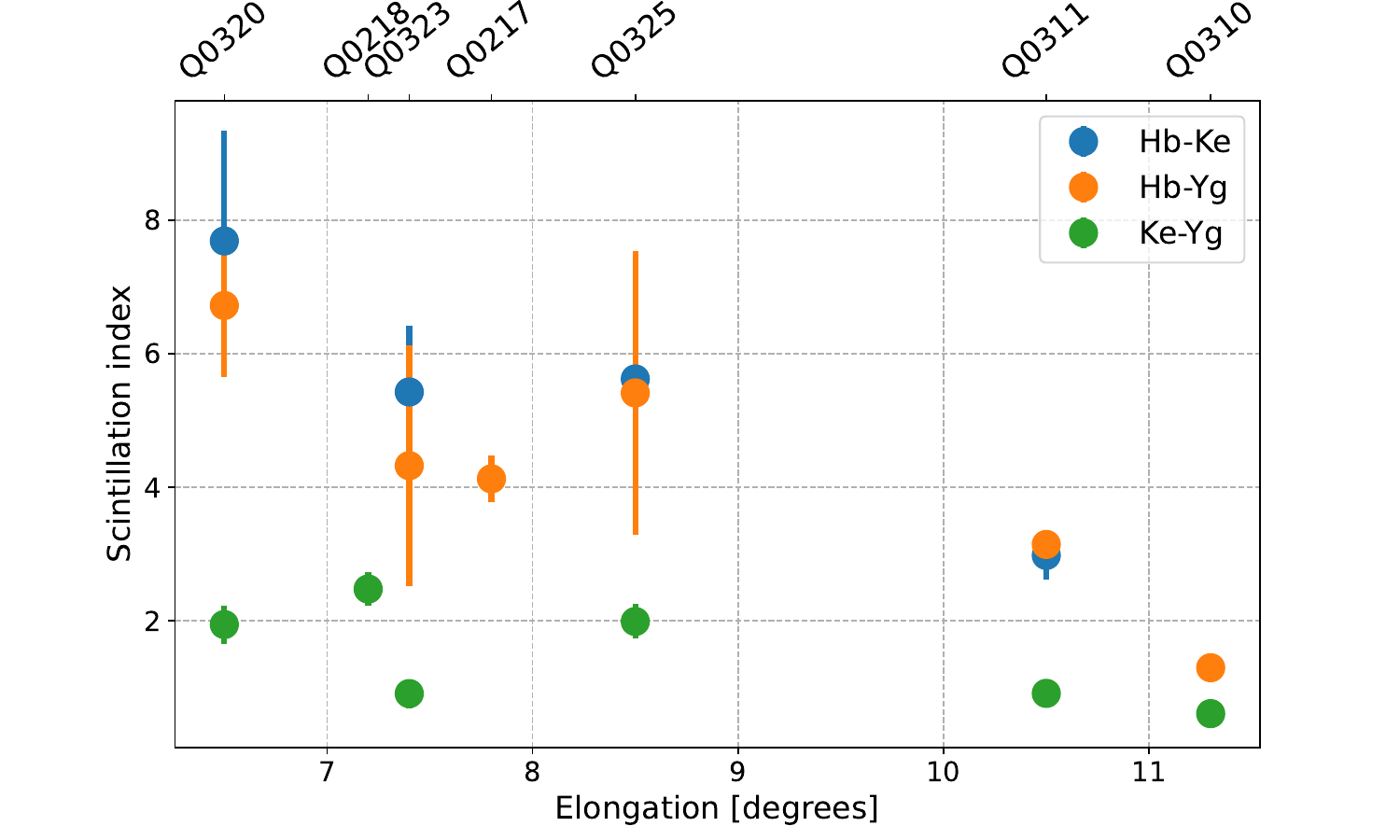}

\caption{The phase scintillation index at the reference frequency of 8 GHz as a function of elongation angle for the three baselines (Hb-Ke, Hb-Yg, and Ke-Yg), with elongation angles ranging from 6.5 to 11.3 degrees. Data points are shown with error bars, where circles represent the mean phase scintillation index across all scans at each elongation angle, and the error bars represent the standard deviation. Missing data for some baselines is due to instrumental or data quality constraints.}
\label{FIG:ipsvselong}
\end{center}
\end{figure}

Figure \ref{FIG:ipsvselong} presents the phase scintillation index at the reference frequency (8 GHz) as a function of the elongation angle for the three baselines: Hb-Ke, Hb-Yg, and Ke-Yg. The elongation angle varies from 6.5 to 11.3 degrees. Each baseline is represented by a different colour for a clear distinction. The data points are plotted with error bars, where the circle represents the mean phase scintillation index of all scans at each elongation angle and the error bars indicate the standard deviation. As a note, the missing data for certain baselines at specific elongation angles are due to instrumental constraints or data quality issues, such as insufficient signal-to-noise ratio or system errors during those observations. These gaps are reflected in the plot by the absence of corresponding data points for those baselines.\\


\begin{figure*}[htbp!]
\newpage
    \begin{subfigure}[b]{0.33\textwidth}
        \includegraphics[width=\textwidth,trim={1cm 0cm 3cm 2cm},clip]{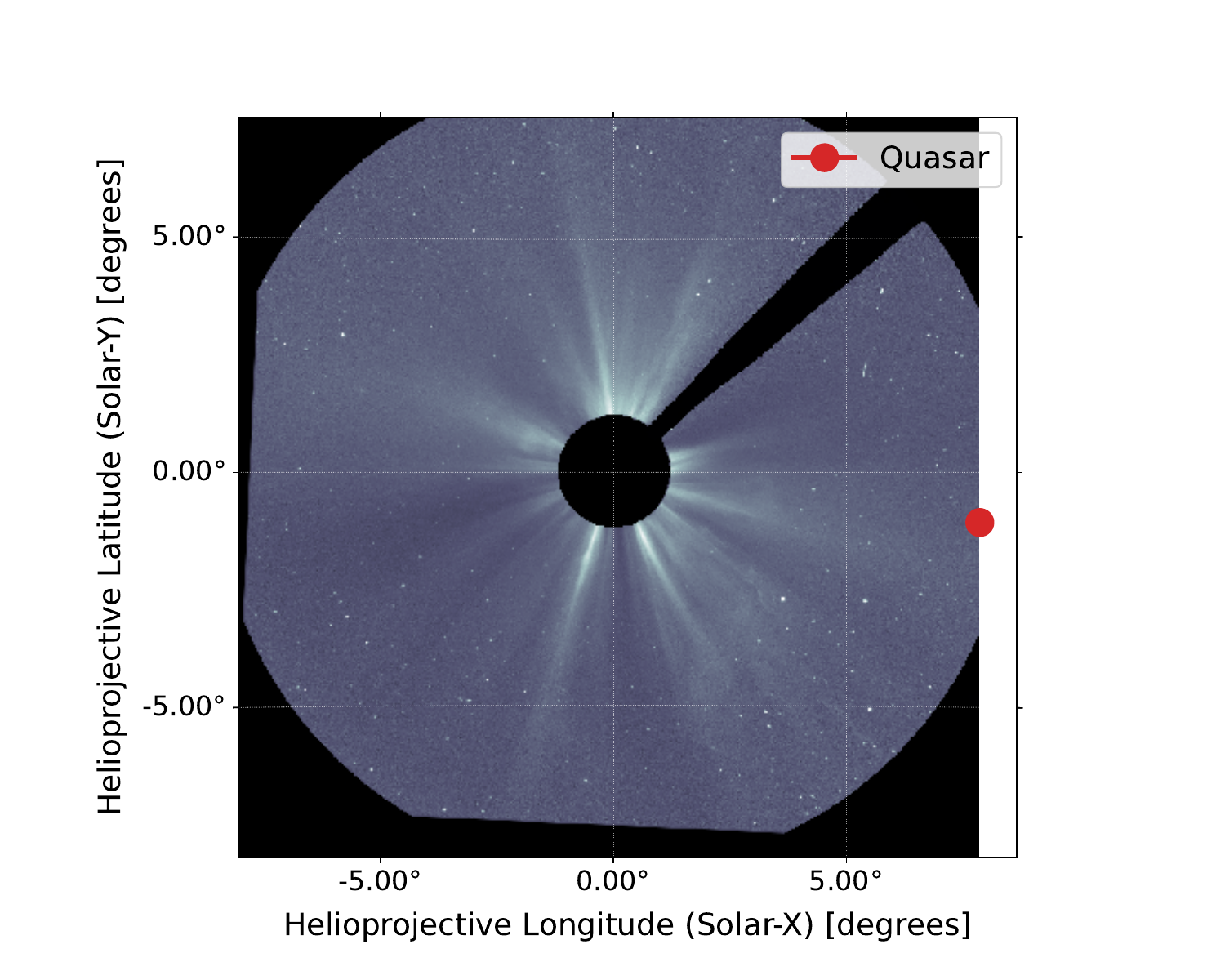}
        \caption{Q0217 - 2024.02.17:0100UT}
        \label{fig:sub01}
    \end{subfigure}
    \begin{subfigure}[b]{0.33\textwidth}
        \includegraphics[width=\textwidth,trim={1cm 0cm 3cm 2cm},clip]{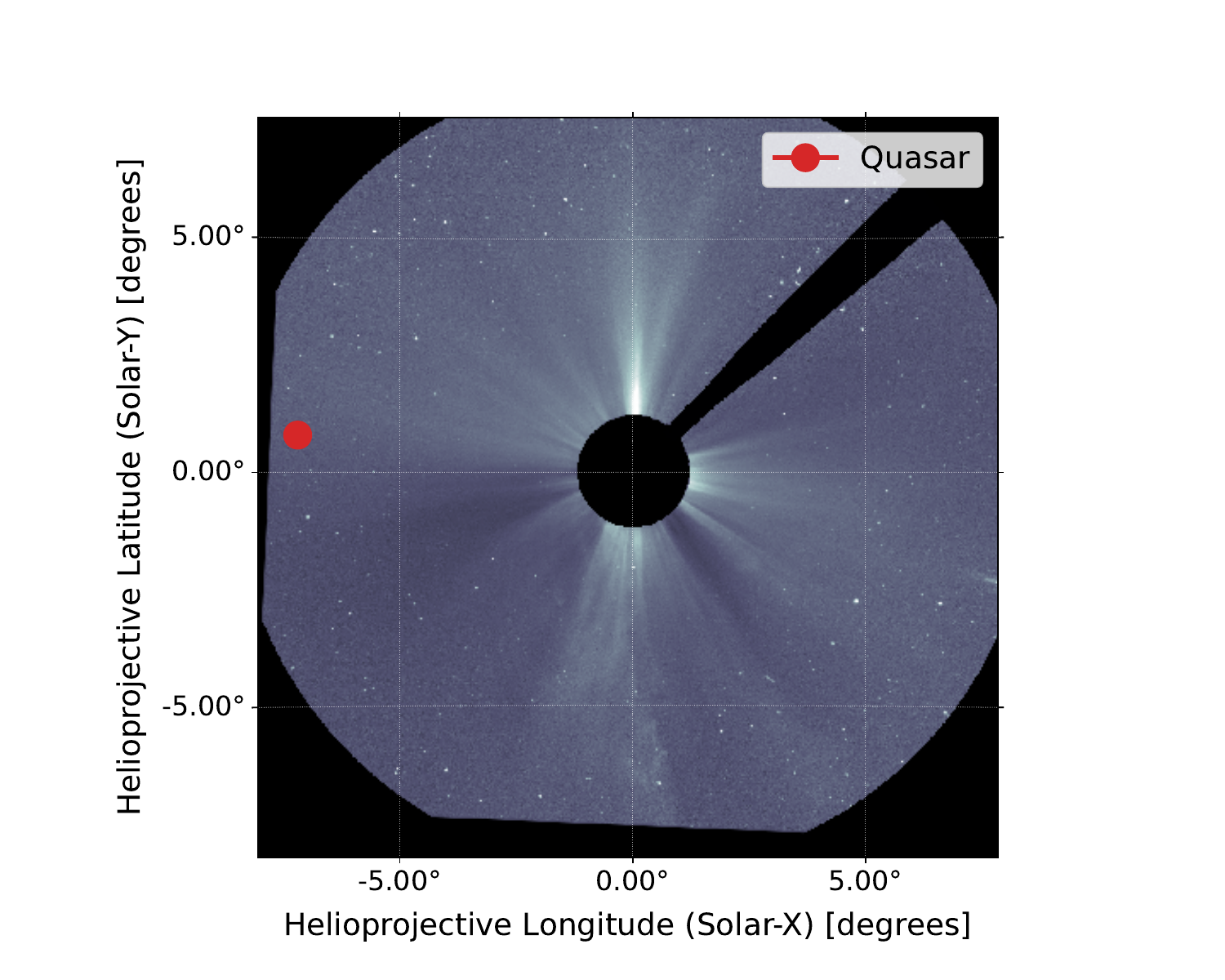}
        \caption{Q0218 - 2024.02.18:0410UT}
        \label{fig:sub02}
    \end{subfigure}
    
    \vspace{0.2cm} 
    
    \begin{subfigure}[b]{0.33\textwidth}
        \includegraphics[width=\textwidth,trim={1cm 0cm 3cm 2cm},clip]{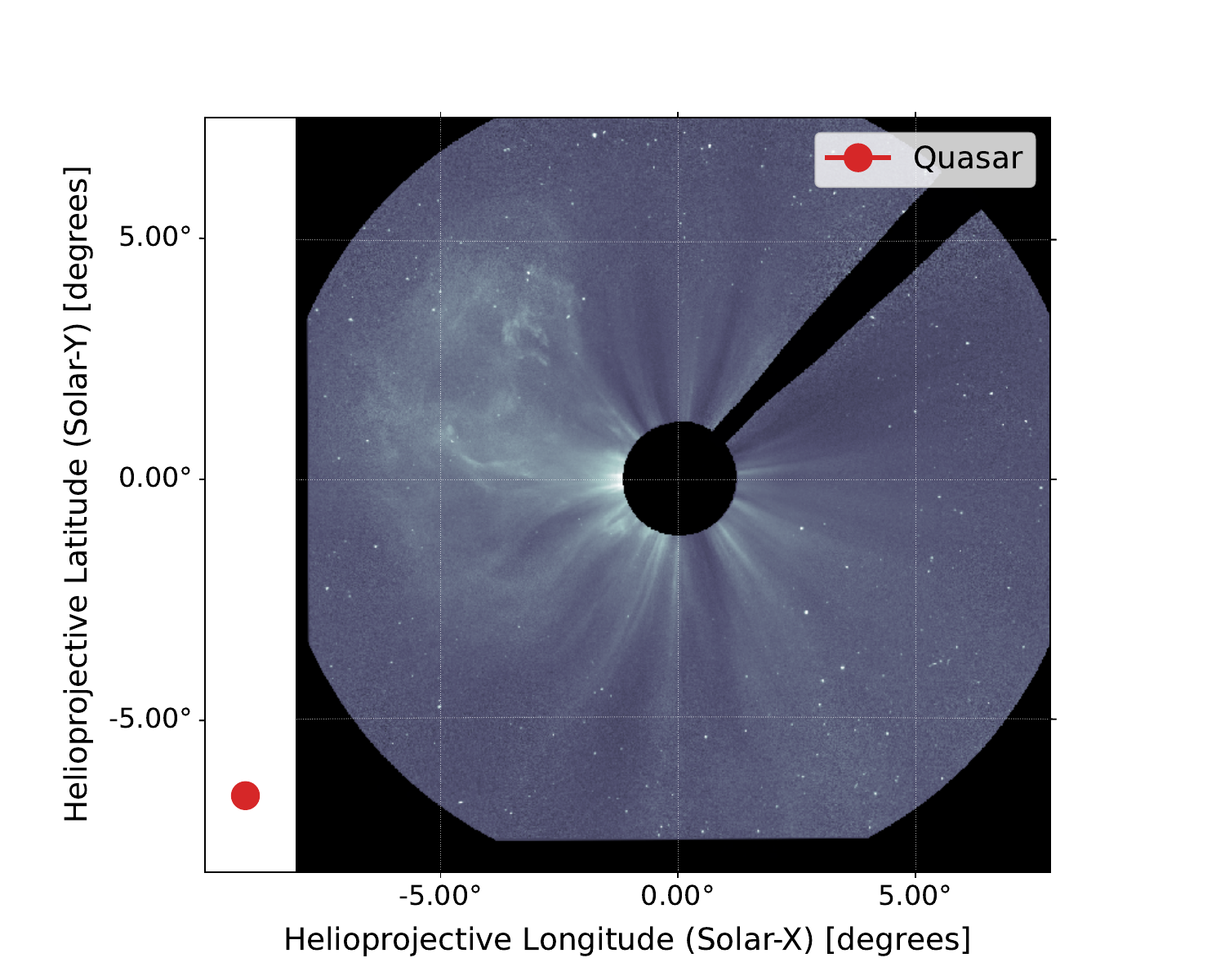}
        \caption{Q0310 - 2024.03.10:0430UT}
        \label{fig:sub1}
    \end{subfigure}
    \begin{subfigure}[b]{0.33\textwidth}
        \includegraphics[width=\textwidth,trim={1cm 0cm 3cm 2cm},clip]{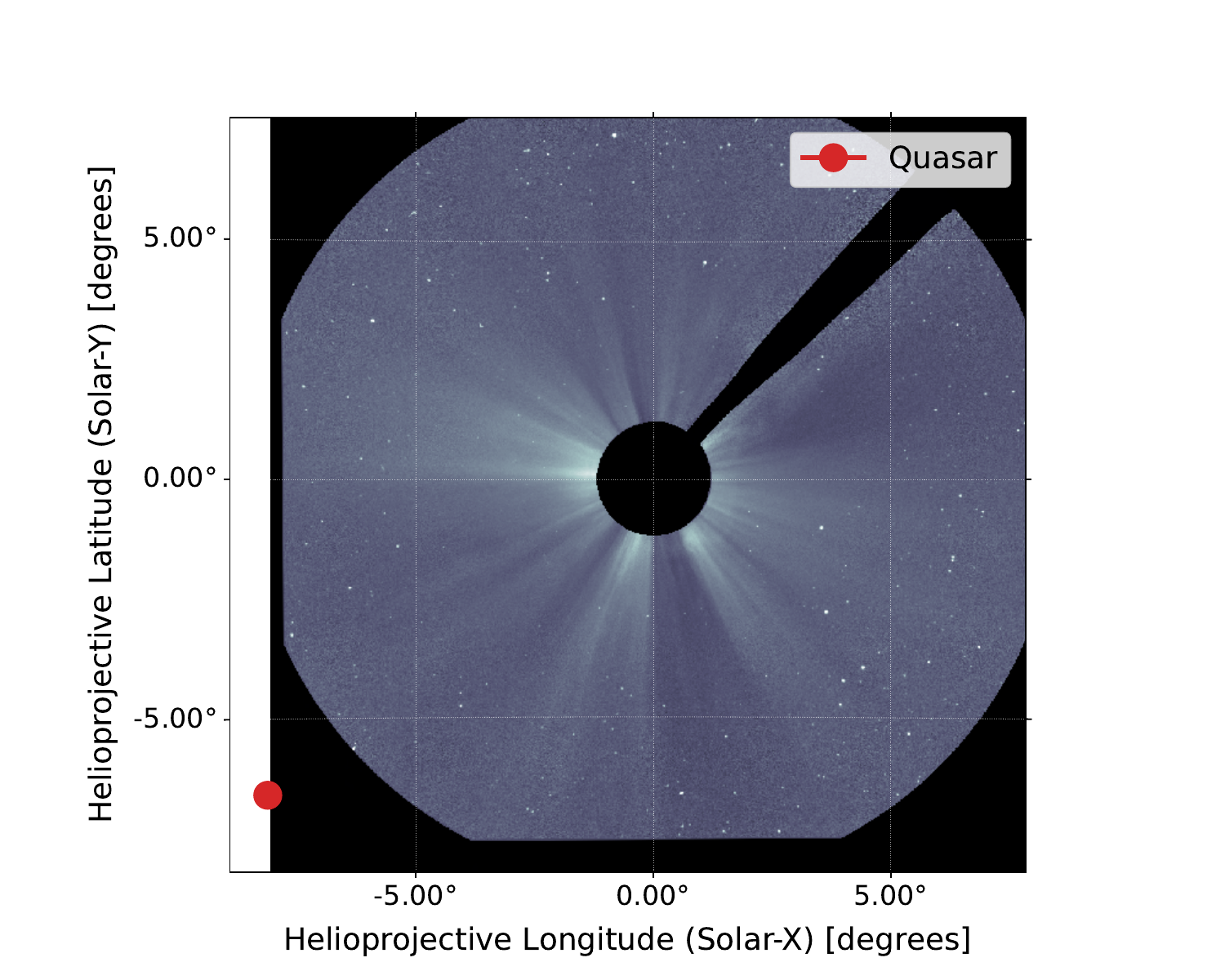}
        \caption{Q0311 - 2024.03.11:0330UT}
        \label{fig:sub2}
    \end{subfigure}
    
    \vspace{0.2cm}

    \begin{subfigure}[b]{0.33\textwidth}
        \includegraphics[width=\textwidth,trim={1cm 0cm 3cm 2cm},clip]{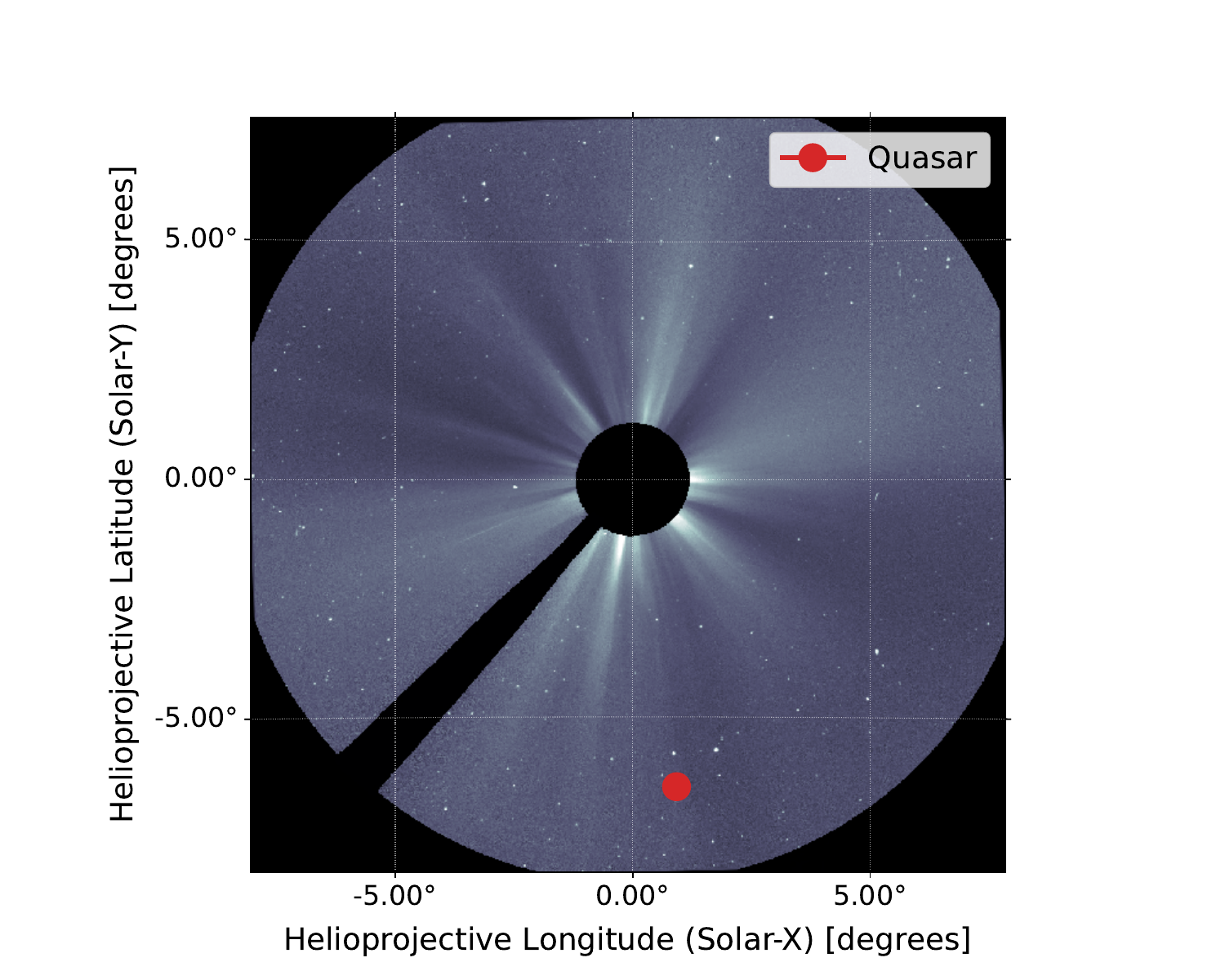}
        \caption{Q0320 - 2024.03.20:0130UT}
        \label{fig:sub3}
    \end{subfigure}
    \begin{subfigure}[b]{0.33\textwidth}
        \includegraphics[width=\textwidth,trim={1cm 0cm 3cm 2cm},clip]{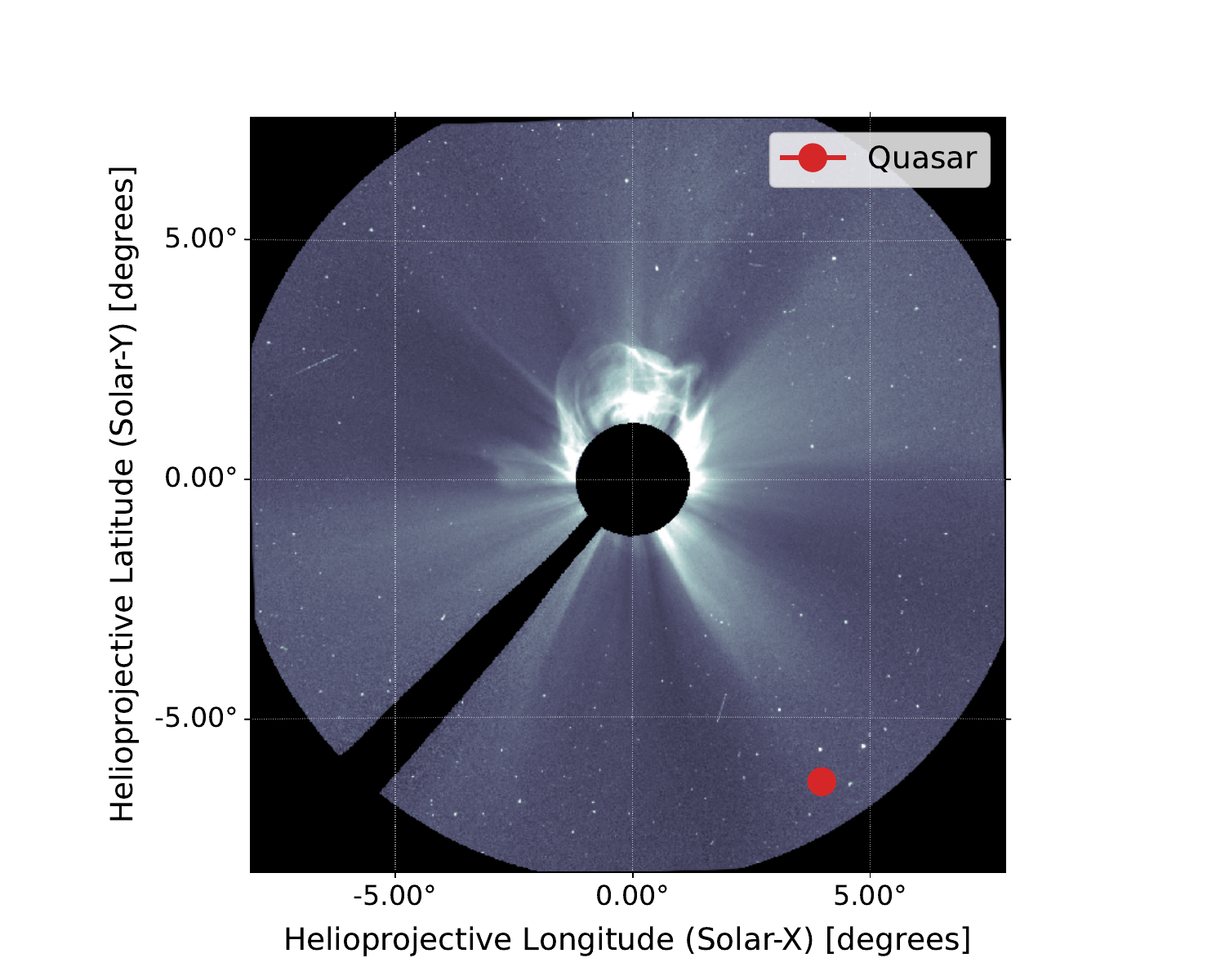}
        \caption{Q0323 - 2024.03.23:0230UT}
        \label{fig:sub4}
    \end{subfigure}

    \vspace{0.2cm}
    
    \begin{subfigure}[b]{0.33\textwidth}
        \includegraphics[width=\textwidth,trim={1cm 0cm 3cm 2cm},clip]{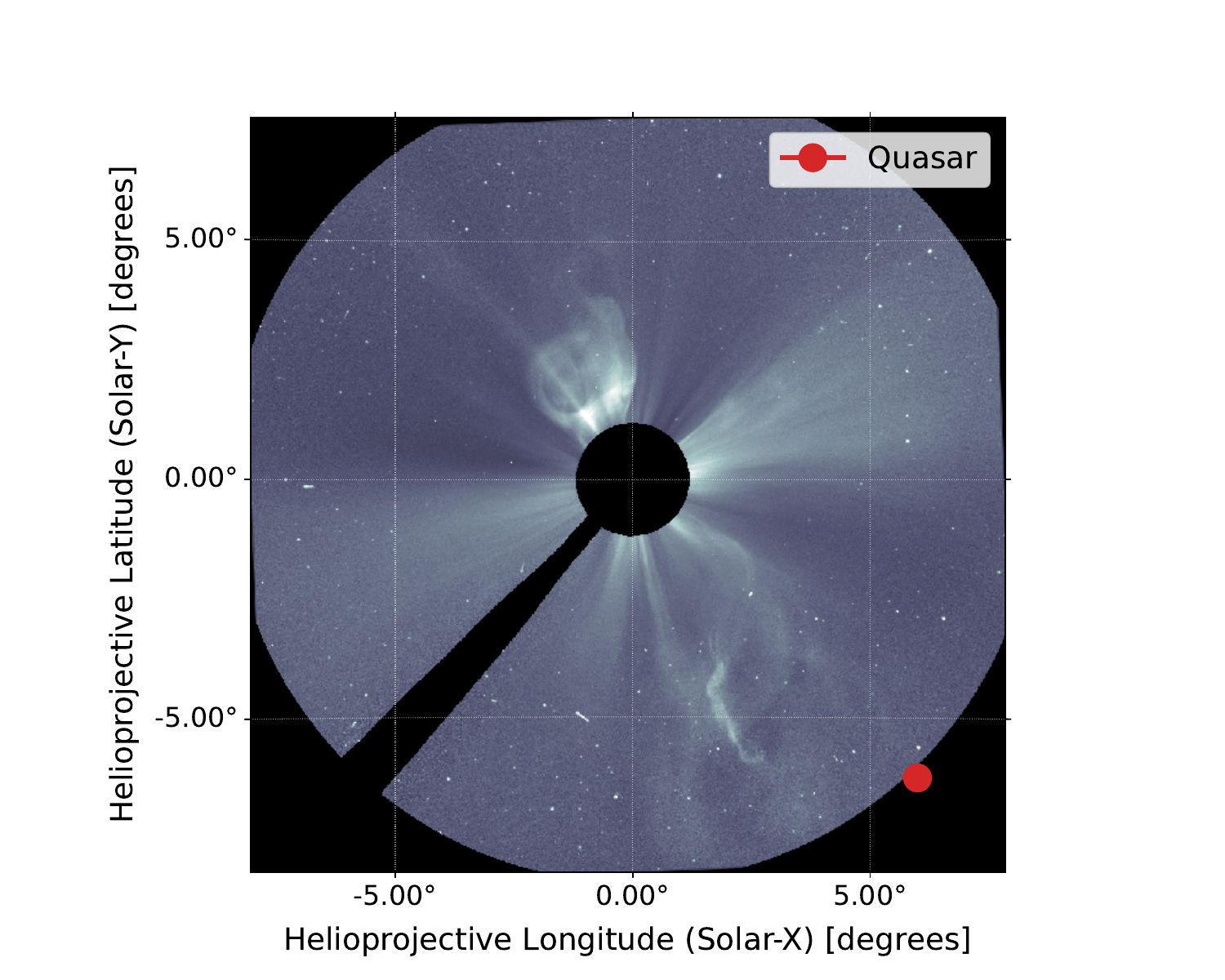}
        \caption{Q0325 - 2024.03.25:0230UT}
        \label{fig:sub5}
    \end{subfigure}

    \caption{LASCO C3 coronagraph images for the observation days. Each image shows the position of the primary source (red circle) relative to the Sun, which is represented by the central disk.}
    \label{fig:coronagraphs}
\end{figure*}


These observations indicated that the phase scintillation indices align well with expected IPS behaviours, confirming theoretical models that predict lower scintillation at larger solar elongation angles \citep{coles1978interplanetary}. However, during session Q0325 the phase scintillation index was higher than expected and exceeded values observed in other sessions with smaller elongation angles.  To understand this anomaly, LASCO C3 coronagraph images were analyzed from all observation days \citep{brueckner1995large}.  Figure \ref{fig:coronagraphs} presents coronagraph images for the observation days. Each image depicts the position of the primary source,  marked by a red circle, relative to the Sun. The image from 25-03-2024 (Figure \ref{fig:sub5}) shows a distinct difference compared to the other days. There appears to be a solar transient passed through the observation line of sight leading to an increase in phase scintillation. In contrast, sessions such as Q0323 (Figure \ref{fig:sub4} ), showed a significant CME  directed towards the North. Since this primary source observed was south of the Sun, this CME is not likely to have directly influenced the line of sight to the quasar. 

\subsection{Phase scintillation index as a function of baseline length}

Figure \ref{FIG:ipsvselong} shows a noticeable difference in the phase scintillation index between different baselines. The magnitude of the phase scintillation index is influenced by factors such as baseline length and orientation, including both radial and vertical components. Specifically, the longest baselines, Hb-Ke and Hb-Yg (3400 km and 3200 km, respectively), exhibit higher scintillation index values. In contrast, the shorter Ke-Yg baseline (2400 km) consistently shows lower scintillation indices. \\

\begin{figure}[htbp!]
\begin{center}
\includegraphics[width=1\columnwidth,trim={1.5cm 0cm 1.5cm 1cm},clip]{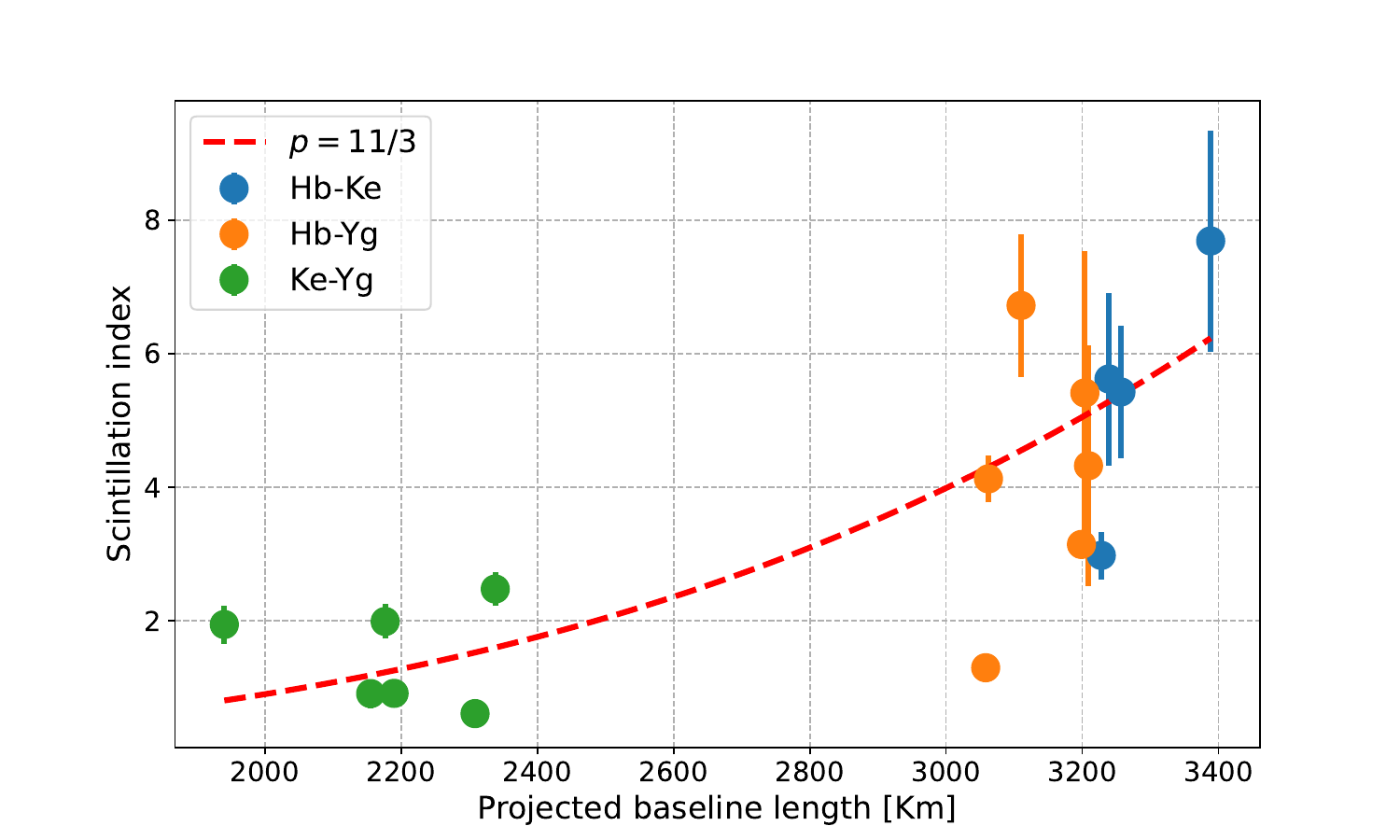}
\caption{Scintillation index as a function of projected baseline length for the Hb-Ke, Hb-Yg, and Ke-Yg baselines. Data points are shown with error bars, where circles indicate the mean phase scintillation index across all scans in each session, and error bars denote the standard deviation. The dashed line represents the Kolmogorov index ($p$) of  11/3. }
\label{FIG:indexvsbl}
\end{center}
\end{figure}

To explore the relationship between the phase scintillation index and the baseline length in more detail, Figure \ref{FIG:indexvsbl} displays the scintillation index against the projected baseline length for the three baselines. The projected baseline length refers to the effective distance between two radio telescopes, projected onto the plane perpendicular to the observed source. This projection determines how much of the baseline length is contributing to the interferometric measurement of the scintillation. \\

Figure \ref{FIG:indexvsbl} shows that as the projected baseline length increases, the scintillation index rises accordingly, suggesting a clear relationship between baseline length and scintillation. The dashed line represents the fitted power law model as a function of projected baseline length, yielding a Kolmogorov index ($p$) of approximately 11/3, consistent with the theoretical expectation from Kolmogorov turbulence and previous studies \citep{coles1989propagation, spangler1995radio}. This agreement between observed data, theory, and previous research underscores the reliability and validity of the measurements, affirming baseline length as a key parameter in studying IPS.\\

\section{Discussion and Conclusion}\label{sec5}

This study demonstrates that broadband Very Long Baseline Interferometry (VLBI) is an effective tool for investigating interplanetary scintillation (IPS) of radio sources near the Sun. Despite the modest size of the 12-meter dishes, the AuScope VLBI array was successfully utilised to measure IPS indices at solar elongation angles from $6.5^\circ$ to $11.3^\circ$. The results showed increased scintillation at lower solar elongations, confirming theoretical models. Expanding to a larger network allows for the observation of multiple lines of sight simultaneously and reconstructs three-dimensional solar wind structures. These findings highlight the utility of broadband VLBI in measuring the solar wind turbulence and dynamics.\\

A key aspect of this study is utilising the broadband system with 32 frequency channels covering 3 to 13 GHz. This wide frequency coverage captures a broad range of turbulence scales and improves the resolution of phase scintillation. The high frequency resolution provided by the 32 channels enhances the separation between signal and noise and characterises the solar wind's frequency-dependent turbulence. This capability detects a wide range of scattering strengths, accommodating both high scattering at lower frequencies and weaker scattering at higher frequencies, thereby refining our understanding of the solar wind’s spectral properties.\\

The new broadband system enhances the capability of VLBI to conduct IPS studies, even with quasars having flux densities as low as 0.5 Jy. By including these weaker radio sources, VLBI provides extensive spatial coverage of the solar wind. This is particularly beneficial for identifying anisotropies and asymmetries, such as those caused by coronal mass ejections (CMEs), thereby contributing to a more detailed analysis of the solar wind's dynamic processes. Additionally, it allows for continuous monitoring of the solar wind’s properties.\\

Moreover, this study shows a clear relationship between the scintillation index and the projected baseline length. As the projected baseline length increases, the scintillation index also rises, demonstrating the direct influence of baseline length on the spatial resolution of IPS measurements. The fitted power law with an index of 11/3, consistent with Kolmogorov turbulence theory, reinforces that the observed scintillation is strongly tied to the solar wind’s turbulent characteristics. Longer projected baselines enable the detection of finer structures within the solar wind. This relationship underscores the importance of baseline length in VLBI studies of solar wind turbulence, enhancing our ability to resolve the intricate details of solar wind fluctuations.\\

The findings from this study support further research using broadband VLBI in IPS studies and its implications for understanding the interplanetary medium. A larger VLBI network will improve spatial coverage and the quality of IPS observations. IPS measurements from VLBI observations of natural radio sources are complementary to those from remote sensing by probes near the Sun \citep{calves2021high} and to those from \textit{in-situ} spacecraft measurements. We aim to combine measurements of spacecraft and geodetic VLBI sources for IPS measurements in the near future. Furthermore,  establishing routine IPS observations as part of geodetic VLBI sessions will support continuous space weather monitoring, offering valuable data for scientific research and operational forecasting, thereby advancing our understanding of solar and space physics.\\

\begin{acknowledgement}
The authors acknowledge Dr. Warren Hankey for his valuable assistance with data transportation.
\end{acknowledgement}

\paragraph{Competing Interests}
None.

\printendnotes
\printbibliography

\end{document}